\documentclass[twocolumn,showkeys,superscriptaddress,preprintnumbers,amsmath,amssymb,showpacs,pre]{revtex4-2}
\usepackage{graphicx}
\usepackage{dcolumn}
\usepackage{bm}
\usepackage{amsmath}
\usepackage{relsize}
\usepackage{wasysym}
\usepackage{color}
\usepackage[colorlinks=true,allcolors=blue,breaklinks=true]{hyperref}
\usepackage{calc}	    	
\usepackage{ifthen}		    
 \usepackage{xspace}

\newcommand{\vcf}{\ensuremath{v_0}}  
\newcommand{\ccfw}{\ensuremath{c_{FW}}}

\newcommand{\dd}{\ensuremath{\mathrm{d}}}
\newcommand{\f}[2]{{
	\ifmmode
		\relax
	\else
		\message{^^JERREUR: ligne \number\inputlineno, la commande 'f' devrait 
			être en mode math^^J^^J}
	\fi
	\ensuremath{%
		\mathchoice%
			{\dfrac{#1}{#2}}
	    	{\dfrac{#1}{#2}}
			{\frac{#1}{#2}}
			{\frac{#1}{#2}}
	}\
}}

\newcommand{\pa}[2][9]{%
	\ifthenelse{#1 = 0}
		{\ensuremath{#2}}{}%
	\ifthenelse{#1 = 1}
		{\ensuremath{(#2)}}{}%
	\ifthenelse{#1 = 2}
		{\ensuremath{\big(#2\big)}}{}%
	\ifthenelse{#1 = 3}
		{\ensuremath{\Big(#2\Big)}}{}%
	\ifthenelse{#1 = 4}
		{\ensuremath{\bigg(#2\bigg)}}{}%
	\ifthenelse{#1 = 5}
		{\ensuremath{\Bigg(#2\Bigg)}}{}%
	\ifthenelse{#1 = 9}
		{\ensuremath{\left(#2\right)}}{}%
}

\newcommand{\Int}[2]{\ensuremath{\mathchoice%
	{{\displaystyle\int_{#1}^{#2}}}
	{{\displaystyle\int_{#1}^{#2}}}
	{\int_{#1}^{#2}}
	{\int_{#1}^{#2}}
	}}

\newcommand{\base}{\ensuremath{\mathcal{B}}\xspace}
\newcommand{\Sum}[2]{\ensuremath{\textstyle{\sum\limits_{#1}^{#2}}}}

\newcommand{\paf}[2]{\ensuremath{\left(\f{#1}{#2}\right)}}

\begin{document}

\title{Dynamic crack growth along heterogeneous planar interfaces: interaction with unidimensional strips}

\author{Aliz{\'e}e Dubois}
\email{alizee.dubois@ens-lyon.fr}
\affiliation{Université Paris-Saclay, CEA, CNRS, SPEC, 91191, Gif-sur-Yvette, France}
\affiliation{ENS Lyon, CNRS, Laboratoire de Physique, UMR 5672, F -69364 Lyon, France}

\author{Daniel Bonamy}
\email{daniel.bonamy@cea.fr}
\affiliation{Université Paris-Saclay, CEA, CNRS, SPEC, 91191, Gif-sur-Yvette, France}

\begin{abstract}
We examine theoretically and numerically fast propagation of a tensile crack along unidimensional strips with periodically evolving toughness. In such dynamic fracture regimes, crack front waves form and transport front disturbances along the crack edge at speed less than the Rayleigh wave speed and depending on the crack speed. In this configuration, standing front waves dictate the spatio-temporal evolution of the local crack front speed, which takes a specific scaling form. Analytical examination of both the short-time and long-time limits of the problem reveals the parameter dependency with strip wavelength, toughness contrast and overall fracture speed. Implications and generalization to unidimensional strips of arbitrary shape are lastly discussed.
\end{abstract}

\date{\today}


\maketitle

\section{Introduction}

Understanding how solids break continues to pose signiﬁcant fundamental challenges. For brittle solids broken under tension, Linear Elastic Fracture Mechanics (LEFM) tackles the difﬁculty by reducing the problem to the destabilization and growth of a dominant preexisting crack (see e.g \cite{Lawn93_book, Bonamy17_crp} for introductions). The theory is based on the fact that, in an elastic material, all dissipative and damaging processes are localized in a small zone around the crack tip, referred to as fracture process zone. Crack destabilization and further motion are then governed by the balance between the ﬂux of mechanical energy released into the fracture process zone from the surrounding material and the dissipation rate in this zone. The former is computable within linear elasticity theory and connects to the stress intensity factor. The dissipation rate is quantiﬁed by the fracture energy. 

LEFM is based on continuum mechanics; as such, it considers homogeneous solids. However, stress concentration at crack tip makes the behavior observed at the continuum-level scale extremely sensitive to small-scale inhomogeneities in the material. Consequences include giant fluctuations in fracture dynamics \cite{Fineberg91_prl,Santucci04_prl,Maloy06_prl,Bares14_prl} and erratic crack paths \cite{Sharon96_prl,Hull99_book,Bouchaud90_epl} incompatible with engineering continuum approach. To capture these features, it has been proposed \cite{Gao89_jam,Schmittbuhl95_prl, Ramanathan97_prl} to consider crack propagation in a solid with spatially-distributed toughness. Within this framework, the onset of fracture is mapped to a critical depinning transition \cite{Bouchaud97_jpcm,Bonamy09_jpd,Ponson09_prl} and crack roughening is interpreted using interface growth models \cite{Barabasi95_book}. This approach succeeds in explaining qualitatively several aspects: self-affine features in fracture roughness \cite{Bouchaud93_prl, Bouchaud97_jpcm, Bonamy11_pr, Ponson16_ijf}, crackling seismic-like dynamics sometimes observed in slowly fracturing systems \cite{Bonamy08_prl,Bares19_pre}, and toughening effects due to microstructural disorder \cite{Roux03_ejma,Patinet13_prl,Demery14_epl}. Yet, this approach was derived within the elastostatic approach, hence, it only applies to slow fracture regimes. 

A major bottleneck is to capture the dynamic stress redistribution through elastic waves, which occurs when a dynamically growing crack interacts with material inhomogeneities. This was first examined \cite{Rice94_jmps, Perrin94_jmps, Benzion95_jmps} in a minimal scalar (antiplanar) model of fracture, and these works show that the crack front will continually roughen under repetitive interactions with heterogeneities. More recently, driven by the availability of analytical solutions for the three dimensional vectorial elastodynamic crack problem \cite{Willis95_jmps, Willis97_jmps}, Ramanathan and Fisher \cite{Ramanathan97b_prl} and Morrissey and Rice \cite{Morrissey98_jmps,Morrissey00_jmps} examined the problem of the crack front interaction with a single asperity. They demonstrate that a new kind of front waves (FW) form and transport perturbations along the crack front without geometric attenuation. Originally invoked for mode I (tensile) loading, Fekak et al. \cite{Fekak20_jmps} recently showed FW also exist for mode II (shear) fracture. Some experiments \cite{Sharon01_nature} suggest FW are responsible for characteristic undulations along fracture surfaces, even  though the precise origin of these undulations remains controversial \cite{Bonamy03_prl,Sharon04_prl,Bonamy04_prl}. These FW were suggested to be the possible source of crack surface roughening in brittle materials \cite{Bouchaud02_jmps,Sharon02_prl,AddaBedia13_prl}. Nevertheless, the processes by which a dynamically growing crack roughens due to interactions with inhomogeneities and the underlying role of FW remain to be clarified. 

We report herein a numerical and theoretical study of dynamic heterogeneous fracture in the simplest configuration: a single propagating crack growing along a two-dimensional (2D) plane made of unidimensional strips parallel to crack propagation with periodically modulated toughness.  The approach is presented in Sec. \ref{Sec1}. We consider first-order perturbations from the situation of straight front propagation at constant speed and apply Morrissey \& Rice's numerical methodology formulating the perturbative solution into the wavenumber-time domain to obtain an equation of motion. The spatio-temporal evolution of local front distortions and local speeds is examined in Sec. \ref{Sec2}. They are shown to result from standing FW, which determines their scaling form.  Section \ref{Sec3} presents an analytical examination of both the short and long time limits of the equation of motion. This provides the dependencies in the parameters concerning strip wavelength, toughness contrast and overall fracture speed. Finally, Sec. \ref{Sec4} demonstrates how the interaction of dynamic crack with unidimensional strips of arbitrary shape can be reconstructed from the superposition of these elementary solutions. 

\section{Theoretical and simulation framework}\label{Sec1}

\subsection{Dynamic crack growth in homogeneous materials: LEFM theory}\label{Sec1.1}

\begin{figure}
\includegraphics[width=\linewidth]{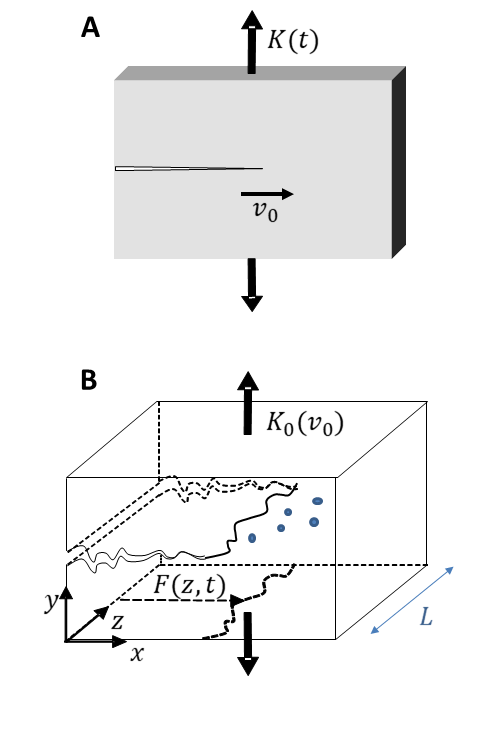}
\caption{Sketch of a crack propagating in a material along with notations used herein. (a) LEFM description of a straight crack front propagating at a speed $v_0$ in a homogeneous brittle solid. The crack front is being loaded in mode I (opening mode). The stress intensity factor is referred to as $K$. (b)  Due to microstructural heterogeneities, the crack front distorts both in-plane and out-of-plane. The approach proposed here neglects out-of-plane roughness. It only considers the first-order in-plane perturbations of the crack front from the reference situation in panel (a) with a straight crack growing at constant speed $\vcf$ due to overall loading $K_0(\vcf)$. The in-plane projection of crack front is parametrized by $F(z,t)=\vcf t +f(z,t)$ where $f(z,t)$ refers to the in-plane perturbations. $L$ is the system size.}
\label{Fig:sketch}
\end{figure}

Let us first consider the situation depicted in Fig. \ref{Fig:sketch}(a) with a crack propagating in a brittle solid loaded in tension (mode I fracture). Herein (and subsequently),  axes $x$, $y$ and $z$ are aligned with the mean direction of crack growth, tensile loading and mean crack front, respectively. $L$ denotes the specimen width along $z$. To address the problem, standard LEFM makes several assumptions:
\begin{itemize}
\item The effect of material inhomogeneities are coarse-grained so that the solid is assumed to be linear elastic, isotropic and homogeneous. The elastic response is fully characterized by Young Modulus $E$ and Poisson ratio $\nu$;
\item The crack is assumed to be straight and perpendicular to tensile loading;
\item The crack front motion is assumed to be invariant along $z$, so that the 3D elastic problem reduced to a 2D plane stress one.   
\end{itemize}  

LEFM then provides the theoretical framework to describe crack growth (see \cite{Freund90_book,Ravichandar04_book} for detailed presentations). We summarize the main steps leading to the equation of motion.

The stress field is shown \cite{freund74_joe} to be singular at the crack tip. It writes:

\begin{equation}
\sigma_{ij}(r,\theta,t) \underset{r \rightarrow 0}{\sim} \frac{K(t)}{\sqrt{2\pi r}}g_{ij}(\theta,\vcf)
\label{Eq:sij}
\end{equation}

\noindent where $(r,\theta)$ are the polar coordinates in the frame $(\vec{e}_x,\vec{e}_y)$ centered at the moving crack tip, and $\vcf$ is the crack speed. The dimensionless functions $g_{ij}(\theta,\vcf)$ providing the angular variations of the stress components are generic (see \cite{Freund90_book,Ravichandar04_book} for details). Conversely, the dynamic stress intensity factor, $K(t)$, depends on both applied loading and specimen geometry at time $t$. It fully characterizes the instantaneous prying force acting on the crack front.

Crack velocity is then predicted from the balance between dynamic energy release rate $G(t)$ and the fracture energy $\Gamma$. $G(t)$ is the flux of mechanical (potential and kinetic) energy released from the specimen as crack propagates over a unit length; $\Gamma$ is the energy dissipated to expose a new unit area of fracture surface. LEFM considers $\Gamma$ a material constant. The difficulty is to determine $G(t)$. To do so, one considers  \cite{Freund90_book,Ravichandar04_book} a region $R$ around the crack tip. This region moves at $\vcf$ with the crack tip. $G$  writes:

\begin{equation}
G = \frac{1}{\vcf}\frac{\dd}{\dd t} \int_R \left(\frac{1}{2}\rho \dot{u}_i^2 + \frac{1}{2}\sigma_{ij} \frac{\partial u_i}{\partial x_j} \right) \dd A,
\label{Eq:Gvss}
\end{equation}

\noindent where $\rho$ is the material density, $u_i$ are the components of the displacement fields, and $\dot{u}_i$ is the time derivative of $u_i$. The summation convention are used for repeated indices. The first and second terms in the integral are the density of kinetic and strain energy, respectively. Invoking Reynolds transport theorem to invert time derivative and spatial integration, then using the equation of momentum conservation $\rho \ddot{u}_i=\partial_j\sigma_{ij}$ and finally applying the divergence theorem, permit to recast Eq. \ref{Eq:Gvss} into: 
  
\begin{equation}
G = \frac{1}{\vcf}\oint_{\partial R} \dot{u}_i \sigma_{ij} n_j + \frac{1}{2}(\rho \dot{u}_i^2 +\sigma_{ij} \frac{\partial u_i}{\partial x_j})\vcf n_x\dd s
\label{Eq:G}
\end{equation}

\noindent where $\partial R$ is the contour of the region $R$ and $n_i$ are the components of the outward normal to the contour. Since we are interested in the mechanical energy flowing into the crack tip, we can make $R$ and $\partial R$ go to zero. Then,  stress field can be replaced by its asymptotic form (Eq. \ref{Eq:sij}). Invoking Hooke's laws for a linear elastic isotropic material, the asymptotic form for the displacement field can also be deduced. It turns out that the contour integral in Eq. \ref{Eq:G} becomes path independent. By choosing the path wisely, it can be shown that:

\begin{equation}
G = \frac{1-\nu^2}{E}A(\vcf)K^2(t),
\label{Eq:Gshort}
\end{equation}

 \noindent where:

\begin{subequations}
\begin{align}
A(\vcf) & = \frac{\vcf^2 \alpha_D(\vcf)}{(1-\nu^2)c_S^2 R(\vcf)},\\ 
\alpha_D(\vcf) & = \sqrt{1-\frac{\vcf^2}{c^2_D}}, \\ 
\alpha_S(\vcf) & = \sqrt{1-\frac{\vcf^2}{c^2_S}}, \\
R(\vcf) & = 4\alpha_D(\vcf)\alpha_S(\vcf) - (1+\alpha_S(\vcf)^2)^2,
\label{chap2F_Gshort}
\end{align}
\end{subequations}

\noindent and $c_D$ is the dilatational wave speed and $c_S$ is the shear wave speed. Finally the equation of motion is obtained by equating $G$ and $\Gamma$, which yields:

\begin{equation}
\frac{1-\nu^2}{E}A(\vcf)K^2(t)  = \Gamma
\label{Eq:LEFM}
\end{equation}

\noindent This equation gives way to a prediction of the crack speed from the knowledge of the dynamic stress intensity factor and the fracture energy.


\subsection{From front distortions to perturbations in dynamic stress intensity factor}

Equation \ref{Eq:LEFM} describes the propagation of dynamic cracks in a homogeneous solid. As in \cite{Ramanathan97b_prl,Morrissey00_jmps}, the effects of microstructural inhomogeneities are now  taken into account by considering a spatially distributed fracture energy:

\begin{equation}
\Gamma(x,y,z) = \Gamma^0(1+\gamma(x,y,z))
\label{Eq:Gammainh}
\end{equation}

\noindent Hence, the crack front is no longer straight. Rather, it exhibits in-plane and out-of-plane distortions [Fig. \ref{Fig:sketch}(b)].  These distortions, in turn, induce variations on the local loading applying along the front.  Willis and Movchan \cite{Willis95_jmps,Willis97_jmps} determined the resulting local variations in terms of stress intensity factors providing several assumptions:

\begin{itemize}  
\item[Hyp. 1] The reference situation is that of a straight front propagating at a constant speed $\vcf$, and the corresponding dynamic stress intensity factor is referred to as $K_0(\vcf)$.
\item[Hyp. 2] In-plane and out-of-plane distortions are small enough such that a first order perturbation analysis is sufficient to relate the local perturbations of stress intensity factor to the front distortions. Note that second order analysis are now available \cite{Willis12_jmps}.  
\end{itemize}  

\noindent As in \cite{Ramanathan97b_prl,Morrissey00_jmps}, we make two additional assumptions:

\begin{itemize}  
\item[Hyp. 3] The out-of-plane distortions are neglected, and the front is assumed to remain within the plane $y=0$. This assumption is justified for quasi-static crack growth since the full three-dimensional description leads to logarithmic out-of-plane roughness \cite{Ramanathan97_prl,Bares14_Frontiers}. However, it should be emphasized that out-of-plane roughness may be rougher in the case of dynamic fracture \cite{Ramanathan97b_prl, Bouchaud02_jmps}.  
\item[Hyp. 4] The terms brought by the non-singular stresses near crack tip in the reference situation ($T$-stress) is neglected. This assumption remains valid as long as $T \leq 0$ (stable in-plane perturbation) and distortion wavelengths are small with respect to the system size and characteristic distances defined by the loading applied \cite{williw2003_iutam}. 
\end{itemize}  

In the following, crack front position at time $t$ is defined by the function $F(z,t)$ [Fig. \ref{Fig:sketch}(b)]. It splits into a global part propagating at $\vcf$ and a distorted part $f(z,t)$ : $F(z,t) = \vcf t + f(z,t)$. Within the hypotheses above, to first order in $f$, the perturbed stress intensity factor at position $z$ writes \cite{Willis95_jmps}: 

\begin{equation}
K(z,t|\vcf)  = K_0(\vcf)\left(1 + Q(z,t|\vcf)  \circledast f(z,t) \right),
\label{Eq:Kperturbed}
\end{equation}

\noindent The operator $ \circledast$ denotes time and space convolution. The function  $Q(z,t|\vcf)$ is provided in \cite{Willis95_jmps} (Eq. 9.14). At this stage, its expression is quite complicated and does not allow for an easy computation.

Introducing Eqs. \ref{Eq:Gammainh} and \ref{Eq:Kperturbed} in Eq. \ref{Eq:LEFM} leads to:

 \begin{equation}		
\frac{1-\nu^2}{E}A\left(\vcf+\frac{\partial f}{\partial t}\right)K_0^2\times(1 + Q  \circledast f)^2 = \Gamma_0(1+\gamma)	
 \end{equation}

\noindent By expanding this equation to first order, and subsequently making $\left( (1-\nu^2)/E \right) A(\vcf) K_0^2= \Gamma_0$ in the obtained equation, we  get:

 \begin{equation}		
\frac{A'(\vcf)}{A(\vcf)} \frac{\partial f}{\partial t} + 2 Q  \circledast f = \gamma(z,x=\vcf t+ f(z,t)),
\label{Eq:Motion_raw_ini}
 \end{equation}

\noindent where $A'(v)$ is the derivative of $A(v)$ with respect to $v$. This equation provides the equation of motion of the crack line in an inhomogeneous landscape of fracture energy. In the following, we make explicit the time dependency of the right-hand term and define $\tilde{\gamma}(z,t)=\gamma(z,x=\vcf t+ f(z,t))$. Hence, equation \ref{Eq:Motion_raw_ini} becomes: 

 \begin{equation}		
\frac{A'(\vcf)}{A(\vcf)} \frac{\partial f}{\partial t} + 2 Q  \circledast f = \tilde{\gamma}(z,t).
\label{Eq:Motion_raw}
 \end{equation}

 \subsection{On the elastodynamic crack kernel: Dispersive relation of crack front waves}

The difficulty at this point is to express the left-hand part of Eq. \ref{Eq:Motion_raw}. As a matter of fact, it conveniently cast in Fourier space $\{k,\omega\}$, such that $\hat{f}(k,\omega) = \int_{-\infty}^{\infty} \int_{-\infty}^{\infty} f(z,t)\exp(-i k z-i \omega t) \dd z \dd t$. Equation \ref{Eq:Motion_raw} writes:

\begin{subequations}
\begin{align}
& - \hat{P}(k,\omega|\vcf) \hat{f}(k,\omega) = \hat{\gamma}(k,\omega) \\
\mathrm{with} \quad & -\hat{P}(k,\omega|\vcf) = 2\hat{Q}(k,\omega|\vcf)-i\omega \frac{A'(\vcf)}{A(\vcf)}
\end{align}
\label{Eq:Motion_fourier}
\end{subequations}

\noindent Here, $\hat{\gamma}(k,\omega)$ is the Fourier transform of $\tilde{\gamma}(z,t)$ and the kernel $\hat{P}(k,\omega|\vcf)$ writes \cite{ramanathan98_prb}:

\begin{equation}
\begin{split}
 \hat{P}(k,\omega|\vcf)  & =  |k| p\left(u=(\omega/k)^2|\vcf\right) \quad \mathrm{with} \\
  p(u|\vcf) & =  \f{2c_R}{c_R^2-\vcf^2}\sqrt{c_R^2-(\vcf^2+u)} \\
               & - \f{c_D}{c_D^2-\vcf^2} \sqrt{c_D^2-(\vcf^2+u)} \\
              &  -\f{1}{\pi}\Int{c_S^2}{c_D^2}\left[\arctan{ \Bigg( 4\frac{\sqrt{1-\eta/c_D^2}\sqrt{\eta/c_S^2-1}}{\pa{2-\eta/c_S^2}^2}}\Bigg) \right. \\
              &   \quad \left.  \times \f{2\vcf^2\eta-\pa{\vcf^2+u}\pa{\eta+\vcf^2}}{\sqrt{\eta \pa{\eta-\vcf^2-u}}\pa{\eta-\vcf^2}^2}\right] \dd \eta \\
\end{split}
\label{Eq:Pkw}
 \end{equation}

\begin{figure}
\includegraphics[width=\linewidth]{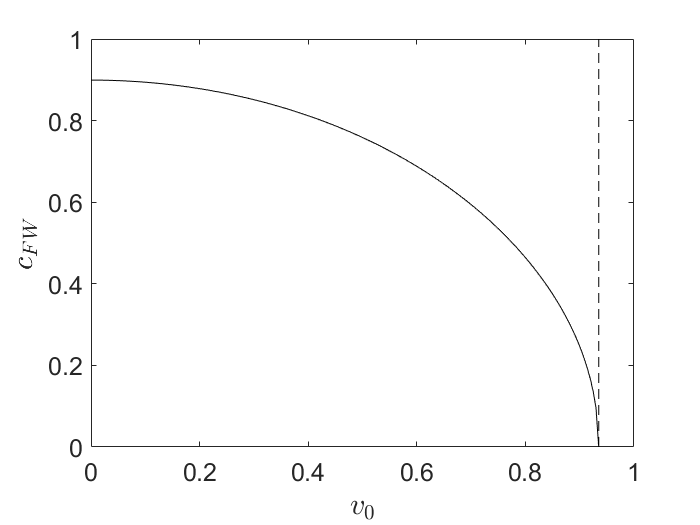}
\caption{Speed of the crack front waves, $\ccfw$, along the front as a function of reference crack speed $\vcf$. Vertical dashed line shows Rayleigh wave speed, $c_R$, which sets the maximum speed $\vcf$ for a mode I crack. All speeds are expressed in $c_S$ units. Poisson ratio is set to $\nu=0.35$, so that  $c_D=2.082 c_S$ and $c_R=0.935 c_S$.}
\label{Fig:c_CFW}
\end{figure}

Interestingly, for all values $\vcf$, there exists a positive finite value $\ccfw(\vcf)=\omega/k$ at which $p(\ccfw^2|\vcf)=0$. Moreover, $\dd \omega/\dd k|_{\omega=\ccfw k}=0$. This is the signature of non-dispersive waves, referred to as crack front waves (FW); any perturbation $f(z,t_0)$ initiated at time $t_0$ subsequently propagates along $z$, both upward and downward and without distortions, at the speed $\ccfw$. The variations of $\ccfw$ with $\vcf$ are plotted in Fig. \ref{Fig:c_CFW}. It starts at a value slightly below $c_R$ for $\vcf=0$, then decreases with $\vcf$ and vanishes as $\vcf = c_R$. This speed is the FW speed measured along $z$, in the frame fixed on the crack line moving at $\vcf$. 

 \subsection{Numerical implementation}

We now turn to the numerical implementation of the problem. The goal is to solve Eq. \ref{Eq:Motion_raw} once a landscape for fracture energy is prescribed. This will reveal the spatio-temporal dynamics of front distortions $f(z,t)$, and subsequently that of the real front $F(z,t)$. Unfortunately, the expression of $Q(z,t)$ is too complex to be handled numerically. Working in the Fourier space in both time and space yields a tractable expression for $\hat{Q}(k,\omega)$ (Eq. \ref{Eq:Pkw}). However, it is difficult to address the quenched disorder term $\gamma(z,x=v_0 t + f(z,t))$ in the equation of propagation. 

Hence, as in \cite{Morrissey00_jmps,ramanathan98_prb}, we choose to work in real space for $t$ and in Fourier space for $z$. This formulation of the equation of propagation gives:  

\begin{subequations}
 \begin{align}
& - \hat{P}(k,t|\vcf) \circledast \hat{f}(k,t) = \hat{\gamma}(k,t), \quad \mathrm{with:}\label{Eq:Motion_kta}\\
 & \{-\hat{P} \circledast \hat{f}\}(k,t) = 2\{\hat{Q}\circledast \hat{f}\}(k,t)+\frac{A'(\vcf)}{A(\vcf)}\frac{\dd \hat{f}}{\dd t}\label{Eq:Motion_ktb}
\end{align}
\label{Eq:Motion_kt}
 \end{subequations}

\noindent Here, $c_R$ refers to the Rayleigh wave speed. The right-hand term of Eq. \ref{Eq:Motion_kta}, $\hat{\gamma}(k,t)$, is the $z$ Fourier transform of $\tilde{\gamma}(z,t)$ After some manipulations available in \cite{Morrissey00_jmps}, this equation becomes:

\begin{equation}
\begin{split}
\frac{\dd \hat{f}}{\dd t} = & -\frac{k^2}{C_v(\vcf)}\int_{-\infty}^t \hat{\base}(
k(t-t')|\vcf) \hat{f}(k,t') \dd t' \\
                                    & \quad + \frac{\hat{\gamma}(k,t)}{C_v(\vcf)} 
\end{split}
\label{Eq:Motion_kt_final}
 \end{equation}
 
\noindent where $C_v(v)$ is:

\begin{equation}		
\begin{split}
C_v(\vcf) =  & \f{c_D}{c_D^2-\vcf^2}- \f{2c_R}{c_R^2-\vcf^2} \\
                  & +\Int{c_S}{c_D}\Theta(\eta) \f{\pa{\eta^2+\vcf^2}}{\pa{\eta^2-\vcf^2}^2}\dd \eta
\end{split}
\label{Eq:Cv}
\end{equation}

\noindent and $\hat{\base}$ is :

\begin{equation}		
\begin{split}
&\hat{\base}(u|\vcf) = c_D\,\f{J_1(\alpha_D c_D u)}{\alpha_D c_D u}-2c_R \f{J_1(\alpha_R c_R u)}{\alpha_R c_R u} \\
& +\f{1}{2} \Int{c_S}{c_D} \Theta(\eta)\left[\f{\pa{\eta^2+v_0^2}}{\pa{\eta^2-v_0^2}}
   J_2(\alpha_\eta \eta u) -J_0(\alpha_\eta \eta u)\right]\dd \eta\\
  \end{split}
\label{Eq:BB}
\end{equation}

\noindent where $J_\nu(u)$ are the Bessel functions of the first kind. The function $\Theta$ involved in Eqs. \ref{Eq:Cv} and \ref{Eq:BB} is: 

\begin{equation}		
\Theta(u) = \frac{2}{\pi}\arctan\left(4\frac{\sqrt{1-u^2/c_D^2}\sqrt{u^2/c_S^2-1}}{(2-u^2/c_S^2)^2}\right)
\label{Eq:Theta}
\end{equation}

\begin{figure}
\includegraphics[width=\linewidth]{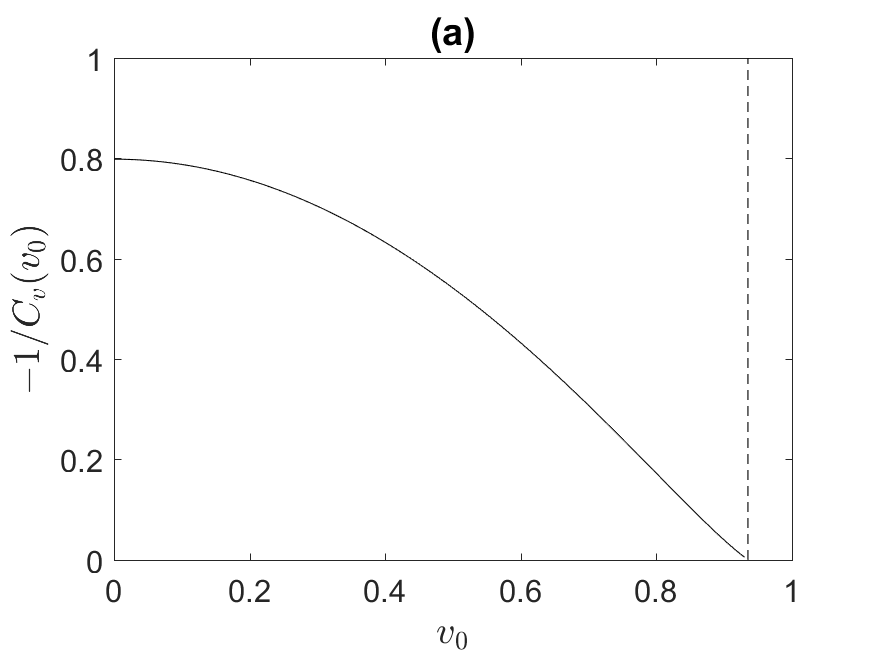}
\includegraphics[width=\linewidth]{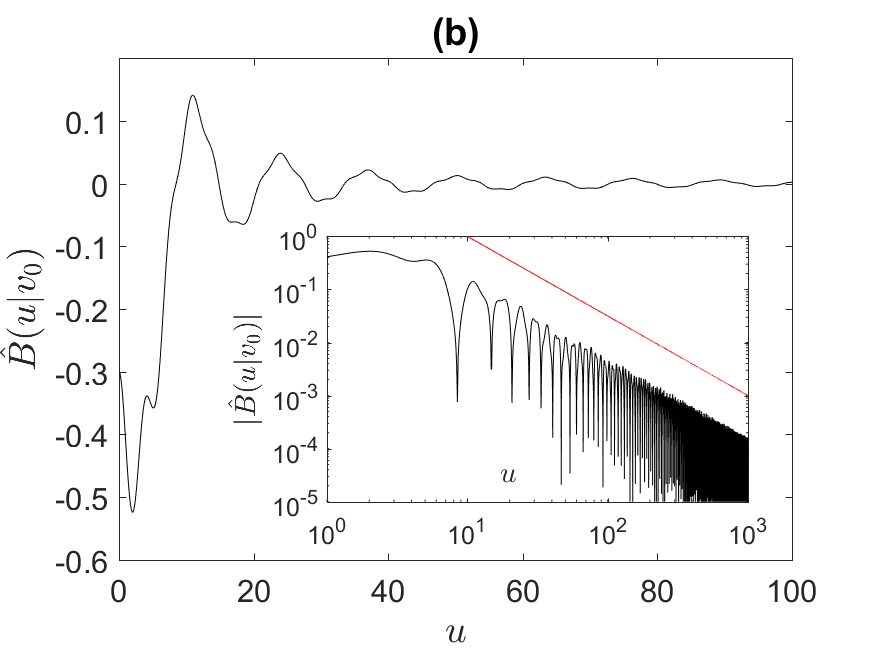}
\caption{Evolution of the two functions involved in the propagation Eq. \ref{Eq:Motion_kt_final}. (a) $-1/C_v$ as a function of $\vcf$. $-1/C_v$ rather than $C_v$ has been plotted since $-1/C_v$ is homogeneous at a speed. Both $-1/C_v$ and $\vcf$ are expressed in $c_S$ unit. Dash vertical line shows Rayleigh wave speed, $c_R$. (b) main panel:  $\hat{\base}(u|\vcf)$ $vs$ $u=k t$ at $\vcf = 0.8 c_S$. (b) inset: same curve in logarithmic scales. Red (straight) line shows a power-law decrease with exponent$-3/2$. All speeds are expressed in $c_S$ units. Poisson ratio is set to $\nu=0.35$, so that  $c_D=2.082 c_S$ and $c_R=0.935 c_S$.}
\label{Fig:cvv_and_BI}
\end{figure}

Figure \ref{Fig:cvv_and_BI}(a)  shows $-1/C_v$ as a function of $\vcf$ ($-1/C_v$ rather than $C_v$ is plotted since $-1/C_v$ has the dimension of speed). $-1/C_v$ starts at a value slightly smaller than $c_R$ at $\vcf = 0$, then decreases with $\vcf$ and vanishes as $\vcf \rightarrow c_R$. This means that faster cracks are less sensitive to inhomogeneities in the fracture energy. A typical curve $\hat{\base}\,\,vs\,\,u$ is shown in Fig. \ref{Fig:cvv_and_BI}(b). $\hat{\base}$ has also the dimension of speed; it is an oscillating function, the amplitude of which decreases with $u$ as $1/u^{3/2}$ in the asymptotic limit [Inset in Fig. \ref{Fig:cvv_and_BI}(b)].

In the following, the crack line $f(z,t)$ is discretized along $z$, $f(z,t)=f_z(t)$ with $z=0,1,2...,L-1$ where $L=1024$ is the system size. Periodic boundary conditions are invoked along $z$, so that $\hat{f}(k,t) = \hat{f}_k(t)$ can be obtained from $f(z,t)$ via discrete Fourier transform. Similarly, the $x$ axis is discretized ($x=0,1,2,...,N-1$) and a discrete map $\gamma_{z,x}=\gamma(z,x)$ is prescribed. At time $t=0$, the front is flat, $f_z(t=0)=0$, and a Euler scheme is used to solve Eq. \ref{Eq:Motion_kt_final} and determine the subsequent time evolution of $f_z(t)$. The different steps involved to move from step $t_i$ to  $t_{i+1} = t_i+\delta t$ are:

\begin{enumerate}
\item Linear interpolation of the map $\gamma_{z,x}$ to get the value at each point $(x=v_0 t_i+f_z(t_i),z)$. This provides $\tilde{\gamma}_z(t_i)=\tilde{\gamma}(z,t_i)$. 
\item Discrete Fourier transform of  $\tilde{\gamma}_z(t_i)$. This provides $\hat{\gamma}_k(t_i)=\hat{\gamma}(k, t_i)$, which is the second right-hand term in Eq. \ref{Eq:Motion_kt_final}.
\item Obtaining the second derivative of $f_z(t_n)$ with respect to $z$  at times $t_n \leq t_i$: $f''_z(t_n) = f_{z+1}(t_n)+f_{z-1}(t_n)-2f_z(t_n)$. Then Discrete Fourier transform of $f''_z(t_n)$ along $z$. 
\item Computation of  $\sum_{n=0}^{n_{max}} \hat{\base}_k (t_n)\hat{f''}_k(t_{i-n})$, where $\hat{\base}_k(t)=\hat{\base}(u=k t|\vcf)$ is obtained using Eq. \ref{Eq:BB}. This provides the first right-hand integral term in Eq. \ref{Eq:Motion_kt_final}. Here, the kernel  $ \hat{\base}_k (t_n)$ was computed prior to applying the Euler scheme and $t_{nmax}$ was chosen so that  $|\hat{\base}(k t)/\hat{\base}(0)|<0.01$ for all $k$ and $t \geq t_{n_{max}}$.
\item Summing the results of steps 2 and 4 to get $\dd \hat{f}_k/ \dd t$ at time $t_i$, and subsequently $\hat{f}_k(t_{i+1})$.
\item Inverse Fourier transform of $\hat{f}_k(t_{i+1})$ to get $f_z(t_{i+1})$.
\end{enumerate}

\noindent The time increment $\delta t$ was chosen so that there is at least 10 points in the smallest period from the Bessel functions involved in $\hat{\base}(u|\vcf)$ (Eq. \ref{Eq:BB}): $\delta t= 2.4/10 \pi c_D\,\alpha_D$. 

In the following, space variables are expressed in pixel size unit ``1'', speeds  are expressed in $c_S$ units, and times are expressed in $1/c_S$ units. The dilatational wave speed is set to $c_D = 2.0817c_S$, leading to a Poisson ratio $\nu = 0.35$. Hence, a Rayleigh wave speed $c_R = 0.935 c_S$ is obtained. Moreover, we will use the notation $f(z,t)$ or $v(z,t)$ rather than $f_z(t)$ or $\dd f_z/ \dd t$ to present and discuss the simulation results. 


\section{Numerical results: dynamic crack growth along sinusoidal strips}\label{Sec2} 

\begin{figure}
\includegraphics[width=\columnwidth]{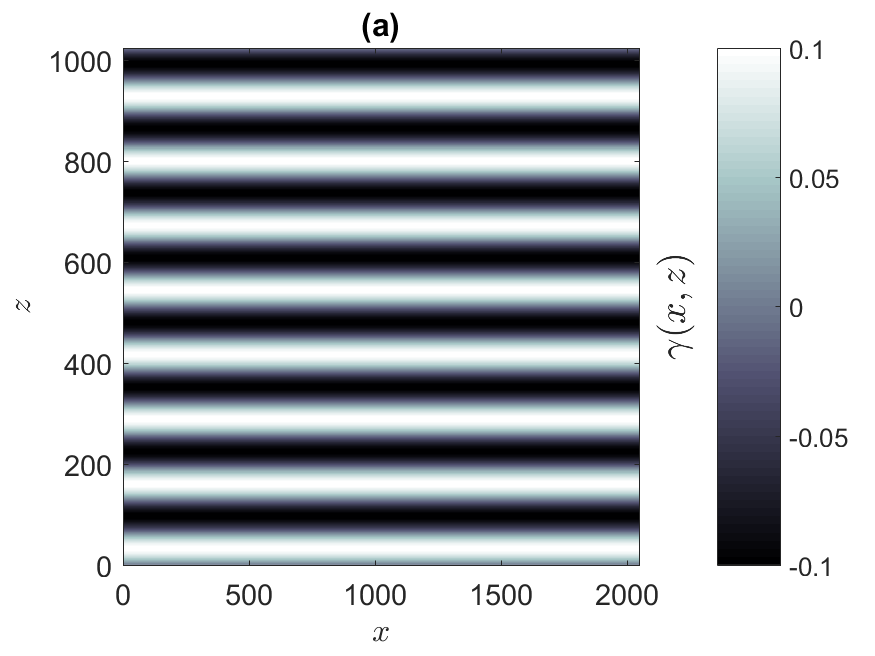}\\
\includegraphics[width=\columnwidth]{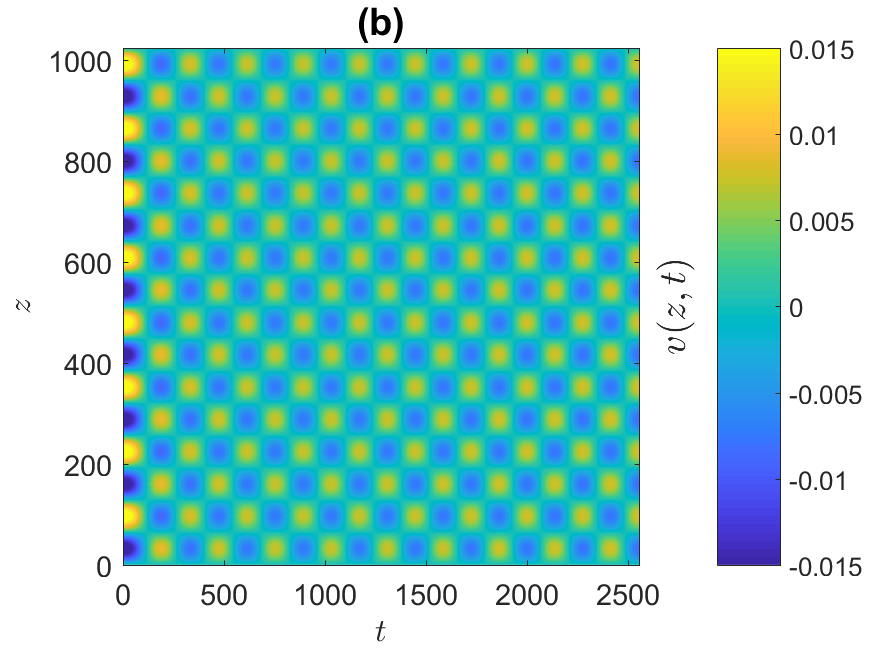}
\includegraphics[width=\columnwidth]{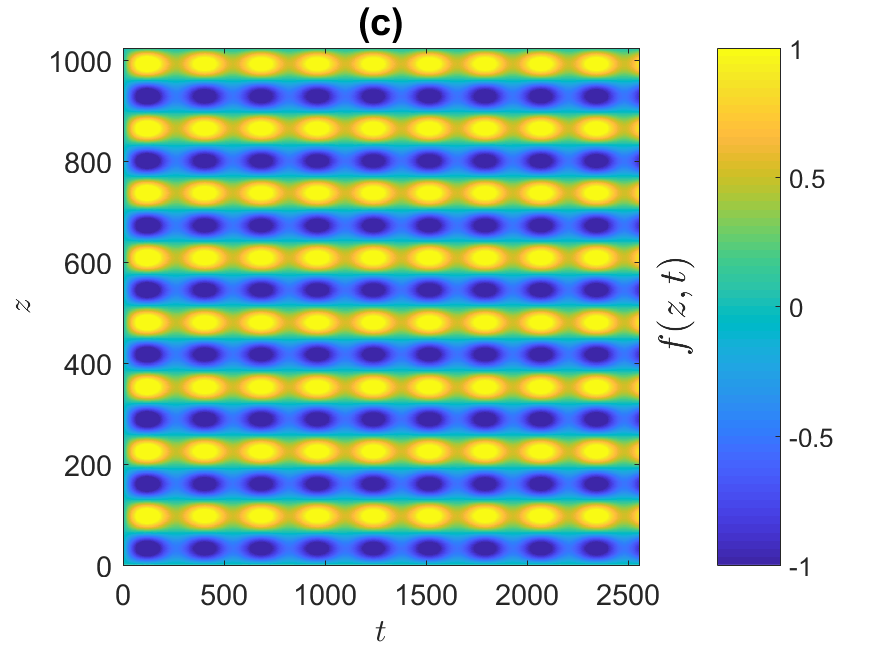}
\caption{(a) Toughness map patterned by parallel bands according to Eq. \ref{Eq:dGz} with $\gamma_0 = 0.1$, $N=1024$ and $\lambda_z = 128$.  (b) Space-time evolution of the local velocity $v(z,t)$ for a crack propagating in the toughness landscape shown in panel (a). (c) Space-time evolution of the distorted part of the front, $f(z,t)$. Poisson ratio is set to $\nu=0.35$, so that  $c_D=2.082 c_S$ and $c_R=0.935 c_S$. In both panels (b) and (c), $\vcf=0.8 c_S$.}
\label{Fig:map_per_z}
\end{figure}

Let us consider the situation depicted in Fig \ref{Fig:map_per_z}(a), with a toughness landscape given by 

\begin{subequations}		
\begin{align}
& \gamma(z,x) = \gamma_0\sin\left(\frac{2\pi z}{\lambda_z}\right), \label{Eq:dGza}\\
i.e. \quad & \tilde{\gamma}(z,t) = \gamma_0\sin\left(\frac{2\pi z}{\lambda_z}\right) H(t)
\label{Eq:dGzb}
\end{align}
\label{Eq:dGz}
\end{subequations}

\noindent  Figure \ref{Fig:map_per_z}(b) displays the local velocity fluctuation $v(z,t)$ of the crack front, and Fig. \ref{Fig:map_per_z}(c) shows the distortion of the crack front $f(z,t)$. These images reveal a checkerboard structure reminiscent of standing waves. This structure results from the interference between the FW propagating upwards and downwards. The form prescribed for $\gamma(z,t)$ in Eq. \ref{Eq:dGza} and the dispersive relation given by the zeros of $p(c|\vcf)$ in Eq. \ref{Eq:Pkw} make it relevant to postulate sine waves for the FW:

\begin{equation}		
A^{\pm}_{FW}(z,t) = \frac{A(t)}{2}\sin \left( \frac{2\pi}{\lambda_z} (z \pm c(t) t)\right)
\label{Eq:sineCFW}
\end{equation}

\noindent where $A(t)$ captures the amplitude variation due to the elastodynamic kernel. Note that in Eq. \ref{Eq:sineCFW}, the FW speed $c(t)$ is also postulated to depend on $t$; as will be seen later, it is only in the long-time limit that $c(t) \rightarrow c_{FW}$. 

Invoking this hypothesis, $v(z,t) = A^{+}_{FW}(z,t) + A^{-}_{FW}(z,t)$. The velocity of the crack front becomes:

\begin{equation}		
v(z,t) = A(t)\sin\left(\frac{2\pi z}{\lambda_z}\right)\cos\left( \frac{2\pi}{\lambda_z}c(t) t\right)
\label{Eq:ST_vansatz}
\end{equation}

\begin{figure}
\includegraphics[width=\linewidth]{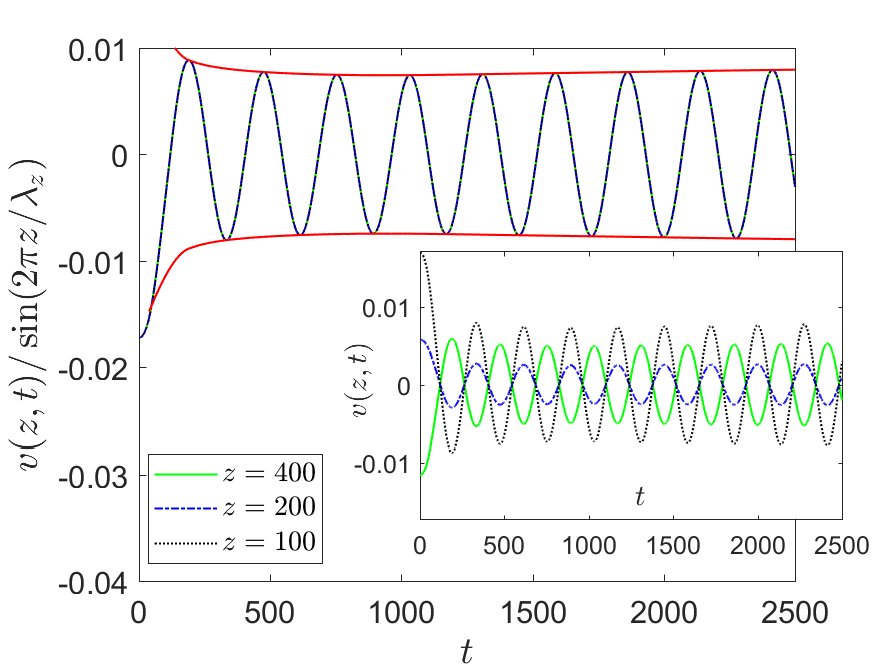}
\caption{Inset: $v(z,t)$ vs. $t$ in a typical simulation at three distinct locations along the crack front: The green (solid oscillating curve), blue (dash-dotted oscillating curve), and black (dotted oscillating curve) correspond to $z=100$ , $z=200$, and $z=400$, respectively. Main panel: $v(z,t)/\sin(2\pi z/\lambda_z)$ vs. $t$ at the same locations. Note the curve collapse,  demonstrating the variable separation proposed in Eq. \ref{Eq:ST_vansatz} for $z$ and $t$. Red (plain non-oscillating) curve shows the amplitude $A(t)$ as obtained by interpolating between the positions and amplitudes of the successive optima. Here, $\vcf=0.8 c_S$, $\gamma_0=0.1$, $N=1024$ and $\lambda_z=128$. Poisson ratio is set to $\nu=0.35$, so that  $c_D=2.082 c_S$ and $c_R=0.935 c_S$.}
\label{Fig:scalingv_per_z}
\end{figure}

\noindent In all our simulations, $v(z,t)$ was found to obey this specific form, irrespective of $v_0$, $\gamma_0$, $\lambda_z$ and $N$ [Fig. \ref{Fig:scalingv_per_z}].

\begin{figure}
\includegraphics[width=\linewidth]{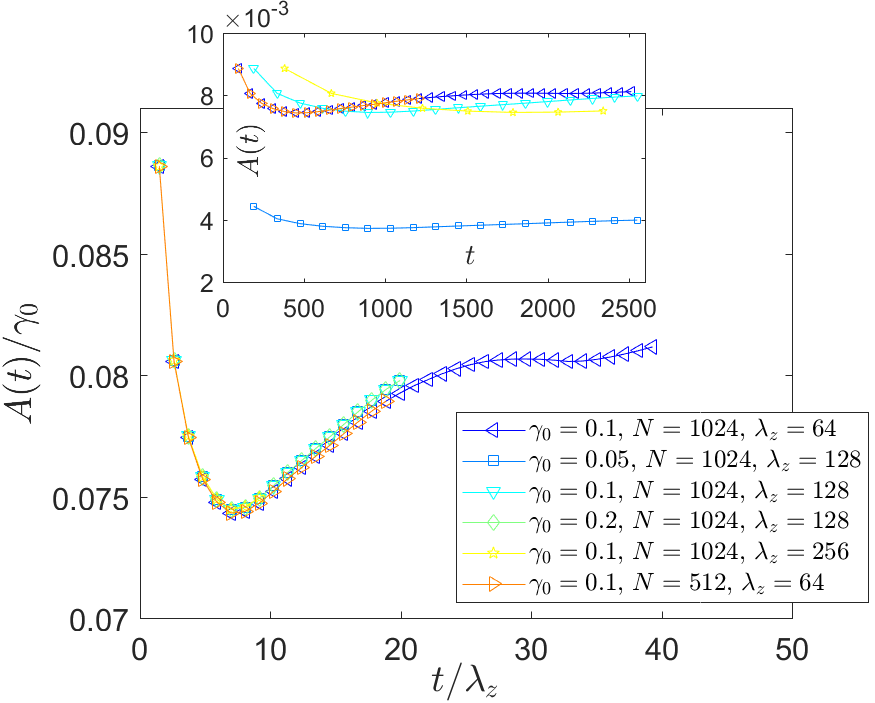}
\caption{Inset: Time evolution of the amplitude, $A(t)$ for $\vcf=0.8 c_S$ and different sets of parameters $\{\gamma_0,N,\lambda_z \}$, the value of which are specified in the legend on the right-hand side of the main panel. Main panel: Collapse using the scaling form proposed in Eq. \ref{Eq:Anorm}. Poisson ratio is set to $\nu=0.35$, so that  $c_D=2.082 c_S$ and $c_R=0.935 c_S$.}
\label{Fig:Aenv_vs_t}
\end{figure}

Let us first consider the function $A(t)$ involved in Eq. \ref{Eq:ST_vansatz}. For each simulation, the waveform amplitude $A(t)$ was obtained by interpolating the successive optima (maxima and minima) of the curve $v(z,t)/\sin(2\pi z/\lambda_z)\, vs.\,t$. Fgure  \ref{Fig:Aenv_vs_t} inset displays the evolution of $A(t)$ with respect to $t$ for fixed $\vcf$ and various set of $\{\gamma_0,N,\lambda_z \}$ parameters. The linearity of Eq. \ref{Eq:Motion_kt_final} with $f$ imposes $A(t)$ to be proportional to $\gamma_0$. Moreover, all curves collapse onto a master curve when time is divided by $\lambda_z$. As a result, we hypothesize the following form for $A(t|v_0,\gamma_0,N,\lambda_z)$: 

\begin{equation}		
A(t|v_0,\gamma_0,N,\lambda_z) = \gamma_0 A^*\left(\left.u=t/\lambda_z\right\vert v_0\right)
\label{Eq:Anorm}
\end{equation} 

\noindent where, for all $\vcf$, $A^*(u)$ exhibits a rapid decrease at small $u$, and subsequently fluctuates around a plateau at large $u$. The analytical analysis detailed in the section\ref{Sec3} characterizes these two regimes.  

\begin{figure}
\includegraphics[width=\linewidth]{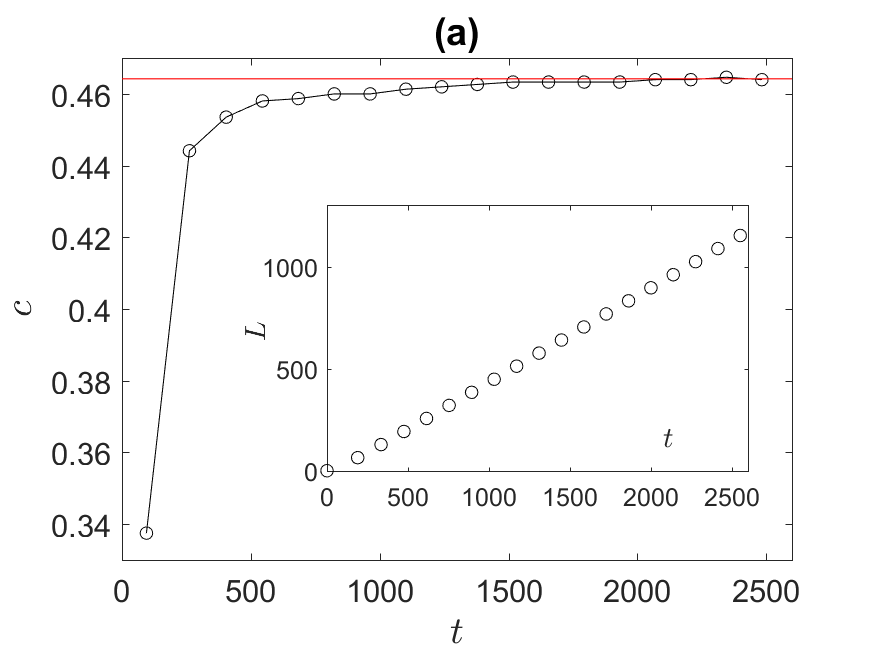}
\includegraphics[width=\linewidth]{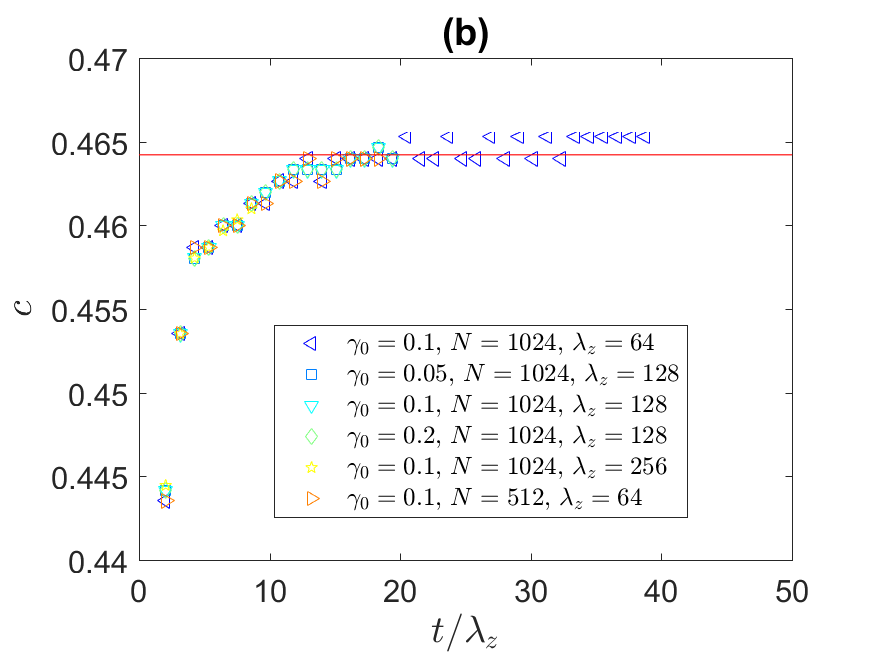}
\caption{FW speed $c(t)$ as measured from the simulations. (a), inset: $L=n \lambda_z/2$ as a function of the time position $t$ of the $n^{th}$ optima (maxima or minima) of the curve $v(z,t)/\sin(2\pi z/\lambda_z)\, vs. \,t$ [Fig. \ref{Fig:scalingv_per_z}].(a), main panel: The discrete differentiation of this $L\, vs.\, t$ curve provides the temporal evolution of $c(t)$. In this panel, $\gamma_0=0.1$, $N=1024$, $\lambda_z=128$ and $\vcf=0.8 c_S $. (b) Collapsed curve $c \, vs. \, t/\lambda_z$ obtained in simulations performed at $\vcf=0.8 c_S$ and different set of parameters $\{\gamma_0,N,\lambda_z \}$, the value of which are specified in the legend on the right-hand side. In both panels (a) and (b), red horizontal line indicates the theoretical value $\ccfw$ so that  $p(\ccfw^2|v_0)=0$, where $p(u|\vcf)$ is provided in Eq. \ref{Eq:Pkw}. All speeds are expressed in $c_S$ units. Poisson ratio is set to $\nu=0.35$, so that  $c_D=2.082 c_S$ and $c_R=0.935 c_S$.}
\label{Fig:c_vs_t}
\end{figure}

Now, let us turn to the form of the FW speed $c(t)$ in Eq. \ref{Eq:ST_vansatz}. To determine the time profile of $c(t)$, we used the following procedure. First, we determined the time positions, $t_{opt}(n)$, of the $n^{th}$ successive optima (maxima and minima) $v(z,t)/\sin(2\pi z/\lambda_z)\, vs\,t$ [Fig. \ref{Fig:scalingv_per_z}]. Second, we plotted $L(n)=n \lambda_z/2$ as a function of $t_{opt}(n)$ [Fig. \ref{Fig:c_vs_t}(a), inset]. Finally, we differentiated the so-obtained curve to get $c(t)$. A typical time profile is plotted in Fig. \ref{Fig:c_vs_t}(a), main panel. The collapse observed in Fig. \ref{Fig:c_vs_t}(b) suggests the following form of $c(t|v_0,\gamma_0,N,\lambda_z)$: 

\begin{equation}		
c(t|v_0,\gamma_0,N,\lambda_z) = c^*(u=t/\lambda_z|v_0).
\label{Eq:cfwnorm}
\end{equation} 


\section{Analytical results}\label{Sec3} 

We now turn to the analytical examination of the equation of motion, in order to discuss the short and long time limits of the front space-time dynamics and the selection of FW speed and amplitude.  

\subsection{Short-time limit}

First, let us examine the short-time limit. To the first order in time, the Bessel functions involved in the definition of $\hat{\base}(u=k t)$ (Eq. \ref{Eq:BB}) writes: $J_0(u)=1 + O(u^2)$, $J_1(u)= u/2 + O(u^2)$ and $J_2(u)=O(u^2)$. As a result, to second order in time, Eq. \ref{Eq:Motion_kt_final} becomes:

\begin{subequations}		
\begin{align}
& \f{\dd \hat{f}}{\dd t}=  -\f{k^2 B_0(\vcf)}{C_v(\vcf)}\int_{-\infty}^t \hat{f}(k,t')\dd t'+\f{\hat{\gamma}(k,t)}{C_v(\vcf)}+O(t^3)\\
& \mathrm{with} \quad  B_0(\vcf)= \frac{1}{2}c_D - c_R -\frac{1}{2}\int_{c_S}^{c_D}\theta(\eta)\dd\eta
\end{align}
\end{subequations}

\noindent By differentiating this equation with respect to time and returning to (z,t) space, we get:

\begin{equation}
\f{\partial ^2f}{\partial t^2}-\f{B_0(\vcf)}{C_v(\vcf)}\f{\partial^2 f}{\partial z^2} =\f{\tilde{\gamma}(z,t)}{C_v(\vcf)} +O(t^2)
\end{equation}

\noindent This partial differential equation is a wave equation of wave speed $c_0 = \sqrt{C_v(\vcf)/B_0(\vcf)}$. This speed is the FW speed at initiation:

\begin{equation}
c(t\rightarrow 0) = c^0_{FW} = \sqrt{\f{C_v(\vcf)}{B_0(\vcf)}}
\label{Eq:c0}
\end{equation}

\begin{figure}
\includegraphics[width=\linewidth]{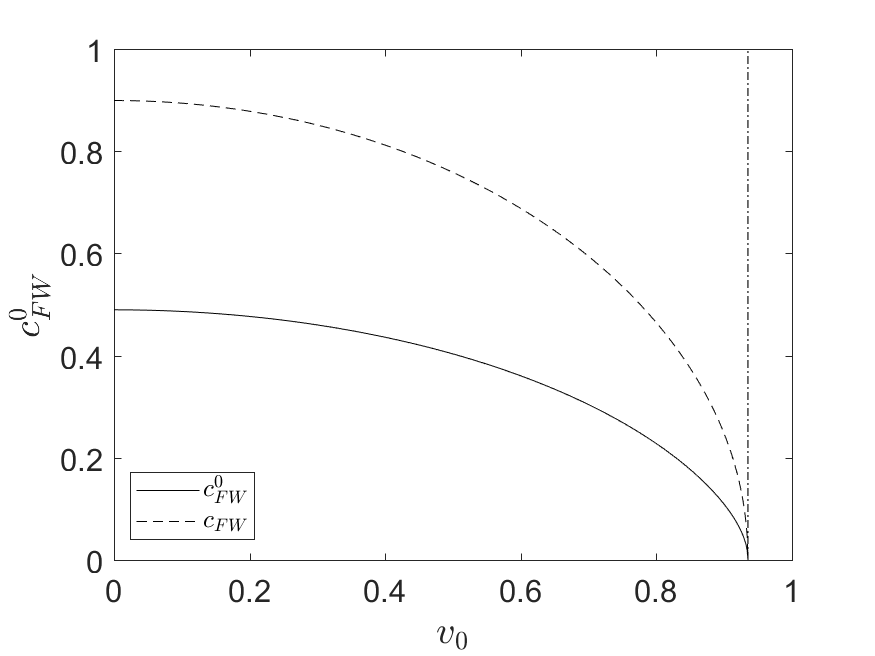}
\caption{FW speed at initiation (solid line), $\ccfw^0=c(t\rightarrow 0)$, as a function of crack speed, $\vcf$. Dashed line recalls the variations of $c_{FW}$ with $\vcf$ [Fig. \ref{Fig:c_CFW}]. Dash-dotted vertical line shows Rayleigh wave speed, $c_R$. All speeds are expressed in $c_S$ units. Poisson ratio is set to $\nu=0.35$, so that  $c_D=2.082 c_S$ and $c_R=0.935 c_S$.}
\label{Fig:c0_CFW}
\end{figure}

Its variations with $\vcf$ are plotted in Fig. \ref{Fig:c0_CFW}. $\ccfw^0(\vcf)$ is about twice smaller than $\ccfw(\vcf)$ and, as $\ccfw$, vanishes as $\vcf \rightarrow c_R$. 

Note finally that $\hat{\base}(u)$ can be expressed as a power series up to any order. By introducing this expression in Eq. \ref{Eq:Motion_kt_final} and subsequently seeking series solutions  $f(k,t)=\sum_mf_m(k)t^m$ of the obtained equation, it is possible to solve analytical Eq. \ref{Eq:Motion_kt_final} up to any order. The whole procedure is detailed in appendix A.   

\subsection{Long time limit}\label{Sec:longtime}

To determine the long-time limit solution, we looked again at the equation of motion in the Fourier space $(k,\omega)$. Making $\hat{\gamma}(k,\omega)=\hat\gamma(k)(\pi\delta(\omega)-i/\omega)$ (where $\pi\delta(\omega)-i/\omega$ is the Fourier transform of $H(t)$ in Eq. \ref{Eq:dGz}) and $\hat{f}(k,\omega)=\hat{v}(k,\omega)/i\omega$ in Eq. \ref{Eq:Motion_fourier} yields in $(k,t)$ space :
 
 \begin{equation}
  \hat{v}(k,t) = -\f{\hat{\gamma}(k)}{2\pi |k|}\int_{-\infty}^\infty \f{1}{p\left(u=(\omega/k)^2\right)}\exp(i\omega t) \dd \omega
\label{Eq:v_kt}
 \end{equation}
 
\noindent The exact computation of the integral is difficult. However, the integrand has two poles: $\omega_+=\ccfw k$ and $\omega_-=-\ccfw k$, which corresponds to the FW solutions that emerge as $t \rightarrow \infty$ (Fig. \ref{Fig:c_vs_t}). Hence, we propose to simplify the integrand and approximate $p(u=(\omega/k)^2)$ by its first order expansion around $u=\ccfw^2$:  

   \begin{equation}
  p_{\infty}\left(u=(\omega/k)^2\right) = p'(c_{FW}^2)\,\pa{\frac{\omega^2}{k^2}-c_{FW}^2}
  \label{Eq:pinfty}
 \end{equation}

\noindent Then, Eq. \ref{Eq:v_kt} becomes:

\begin{equation}
  \hat{v}_\infty(k,t) = -\f{\hat{\gamma}(k)|k|}{2\pi p'(c_{FW}^2)}\int_{-\infty}^\infty \f{\exp(i\omega t)}{(\omega^2-c_{FW}^2 k^2)} \dd \omega
\label{Eq:v_kt_infty}
 \end{equation}

The integral evaluation is detailed in Appendix B. It gives:

\begin{equation}
v_{\infty}(z,t)=\f{\gamma_0}{\ccfw p'(\ccfw^2)}\sin\paf{2\pi z}{\lambda_z}\sin\pa{\f{2\pi\ccfw t}{\lambda_z}}
\label{Eq:vssfinal}
\end{equation}

\begin{figure}
\includegraphics[width=\linewidth]{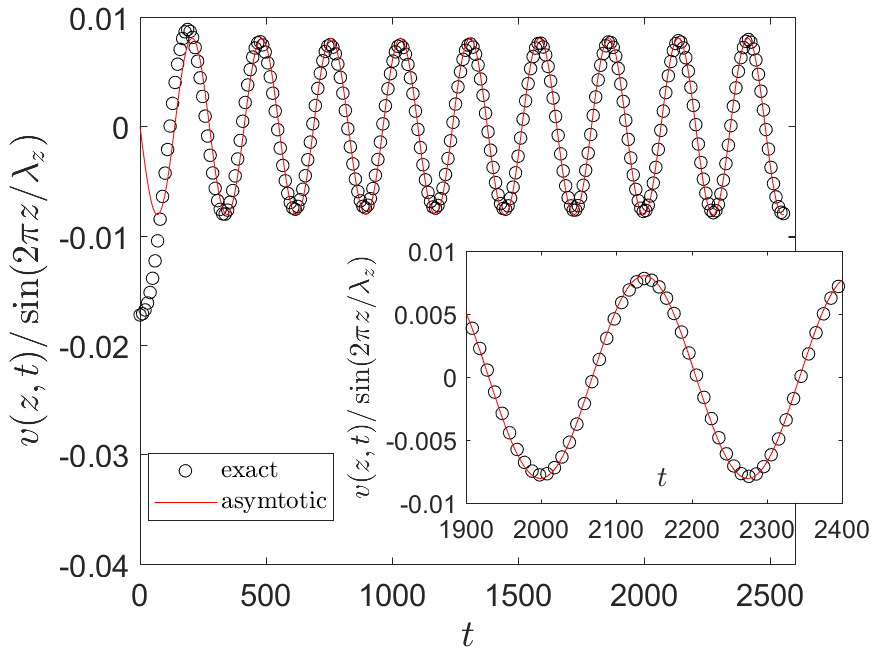}
\caption{Main panel: Comparison between $v(z,t)$ as obtained by simulations (black circles) and $v_\infty(z,t)$ as predicted analytically by Eq. \ref{Eq:vssfinal} in the long-time limit regime (red line). Inset: zoom on a smaller time scale to emphasize the overlap. Here, $\vcf=0.8$, $L=1024$, $\gamma_0=0.1$ and $\lambda_z=0.1$. Poisson ratio is set to $\nu=0.35$, so that $c_D=2.082$ and $c_R=0.935$. All speeds are expressed in $c_S$ unit.}
\label{Fig:asympt-v_vs_t}
\end{figure}

\noindent As shown in Fig. \ref{Fig:asympt-v_vs_t}, this solution superposes almost exactly the numerical solution in the long time limit, as soon as $t\gg\lambda_z/c_{FW}$. This justifies {\em a posteriori} the approximation proposed in Eq. \ref{Eq:pinfty}.

The long-time limit value of normalized amplitude $A^*$ [Eq. \ref{Eq:Anorm}] can then be deduced:

\begin{equation}
A^*(t\gg\frac{\lambda_z}{c_{FW}})=A^*_{\infty}=\f{1}{\ccfw |p'(\ccfw^2)|}
\label{Eq:Ass}
\end{equation}

\begin{figure}
\includegraphics[width=\linewidth]{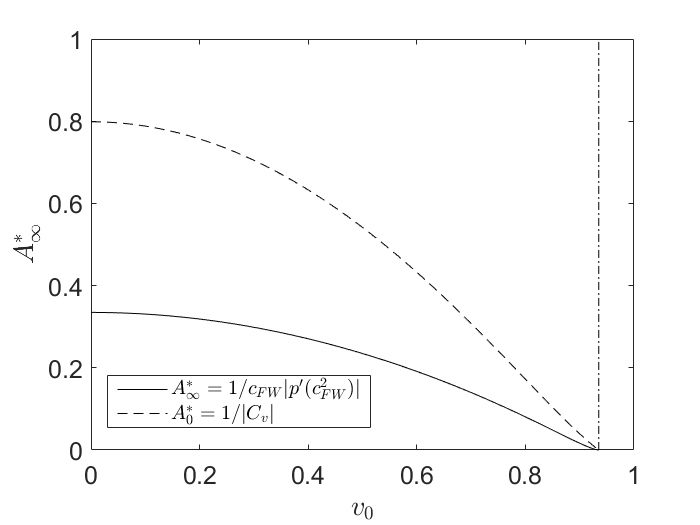}
\caption{Full line: FW normalized amplitude $A^*_{\infty}=A(t/\lambda_z>>1)/\gamma_0$ as a function of crack speed, $\vcf$. Dash line recalls the variations of $A^*(t=0=-1/C_v(v_0))$ with $\vcf$ [Fig. \ref{Fig:cvv_and_BI}(a)].Here, $\ccfw^0$, $\ccfw$ and $\vcf$ are normalized by $c_S$. Poisson ratio is set to $\nu=0.35$, so that $c_D=2.082$ and $c_R=0.935$. All speeds are expressed in $c_S$ unit}
\label{Fig:Ass}
\end{figure}

\noindent Its variations with $\vcf$ are plotted in Fig. \ref{Fig:Ass}. Its value is about twice smaller than the FW amplitude at initiation, $A^*_0=1/|C_v(\vcf)|$. As $A^*_0$, $A^*_{\infty}$ vanishes as $\vcf \rightarrow c_R$.

To interpret this long time behavior, it is of interest to replace $p(u)$ by its approximation $p_\infty$ in Eq. \ref{Eq:Pkw} and express the resulting equation of motion back in the direct $(z,t)$ space. It gives: 
 
   \begin{equation}
  \frac{1}{c_{FW}^2}\f{\partial ^2f_\infty}{\partial t^2}-\,\f{\partial^2 f_\infty}{\partial z^2}= S(z,t),
  \label{Eq:wave_infty}
 \end{equation}
 
 \noindent where $S(z,t)$ is a function of $\gamma(z,t)$:
 
    \begin{equation}
  S(z,t)=\f{1}{(2\pi)^2}\int_{-\infty}^{\infty} \int_{-\infty}^{\infty}  \frac{|k|\,\hat{\gamma}(k,\omega) }{c_{FW}^2\,p'(c_{FW}^2)}e^{(i \omega t+ ikz)} \dd k \dd \omega
  \label{Eq:source_infty}
 \end{equation}
 
 \noindent In other words, the long-time limit solution $f_\infty(z,t)$ is the solution a simple 1D wave equation (Eq. \ref{Eq:wave_infty}) with a source term given by Eq. \ref{Eq:source_infty}. 
 
\subsection{Approximate solution over whole time range}

\begin{figure}
\includegraphics[width=\linewidth]{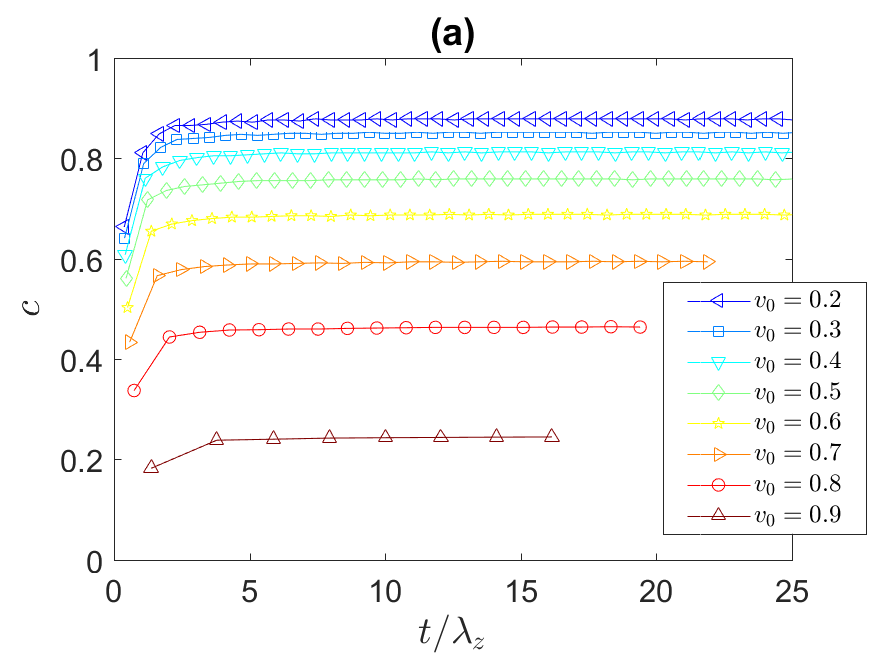}
\includegraphics[width=\linewidth]{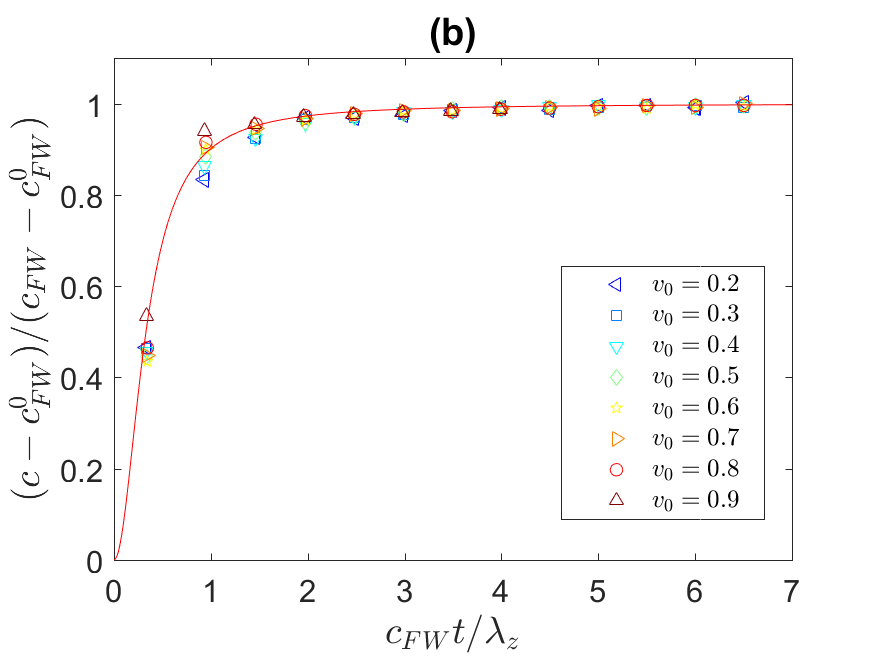}
\caption{(a) (a) FW speed, $c(t)$, as a function of $t/\lambda_z$ at different $\vcf$. Both $c(t)$ and $\vcf$ are expressed in $c_S$ units. (b) Curve collapse obtained using Eq. \ref{Eq:cfull}. The different data point symbols correspond to different values $\vcf$ as specified in the legend on the right-hand side of both panels (a) and (b). Red solid line in panel (b) is a fit $g_c(u)=(a u)^2/(1+(au)^2)$ with a fitted parameter $a=2.78\pm 0.05$ ($\pm$ indicates a $95\%$ confident interval). Poisson ratio is set to $\nu=0.35$, so that $c_D=2.082 c_S$ and $c_R=0.935 c_S$. }
\label{Fig:cfull}
\end{figure}

Let us now use the analytical results obtained in sections IV.A and IV.B to interpret the time profiles of the FW speed $c$ and the amplitude $A$ entering into Eq. \ref{Eq:ST_vansatz}.  Figure \ref{Fig:cfull}(a)  shows the curves $c\:vs.\: t/\lambda_z$ obtained at various fracture speeds $\vcf$. All curves can be superimposed by making $c \rightarrow (c-\ccfw^0)/(\ccfw-\ccfw^0)$ and $t \rightarrow \ccfw t /\lambda_z$ [Fig. \ref{Fig:cfull}(b)] and the resulting curve is well approximated by:

\begin{equation}
\begin{split}
& \f{c - \ccfw^0}{\ccfw - \ccfw^0} = g_c\left( u=\f{\ccfw t}{\lambda_z} \right),\\
& \mathrm{with} \quad g_c(u)\approx \f{(a u)^2}{1+(a u)^2},
\end{split}
\label{Eq:cfull}
\end{equation}

\noindent where $a$ is a fitting parameter: $a \approx 2.8$.

\begin{figure}
\includegraphics[width=\linewidth]{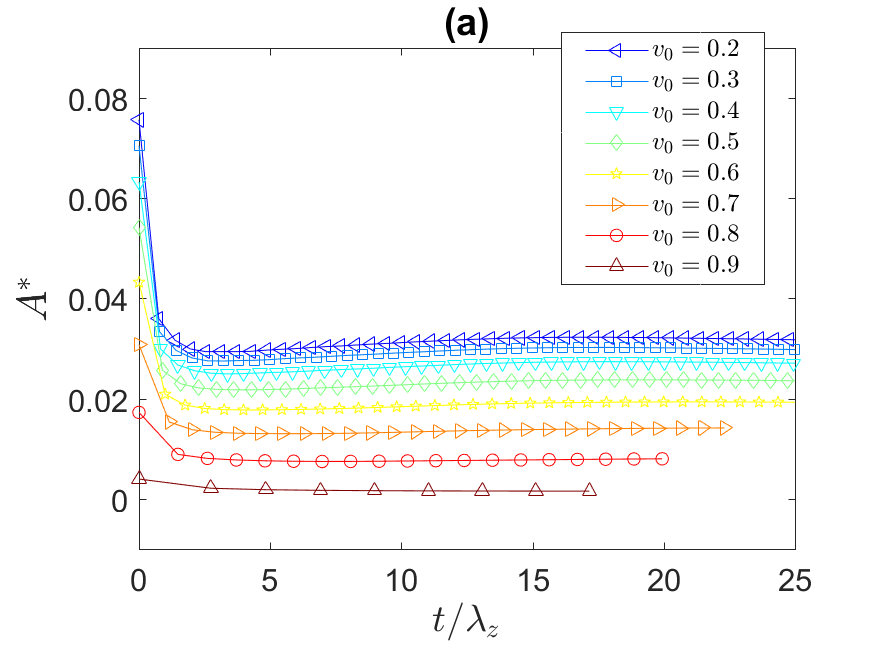}
\includegraphics[width=\linewidth]{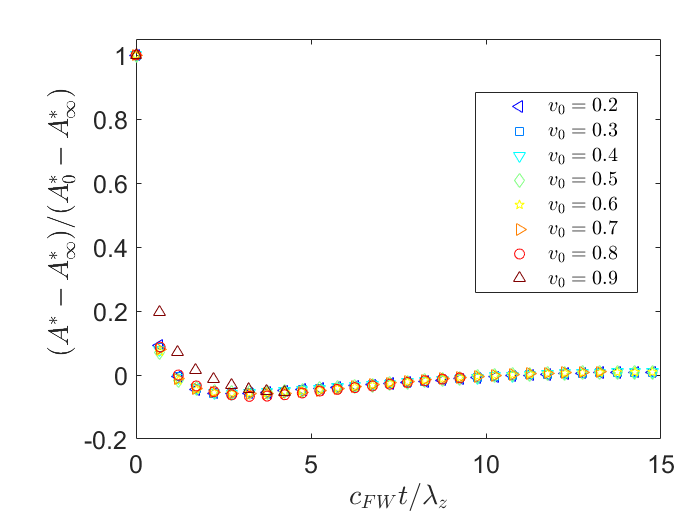}
\caption{(a) normalized FW amplitude, $A^*(t)$, as a function of $t/\lambda_z$ at different $\vcf$. Both $A^*(t)$ and $\vcf$ are expressed in $c_S$ units (b) Curve collapse obtained using Eq. \ref{Eq:Afull}. The different data point symbols correspond to different values $\vcf$ as specified in the legend on the right-hand side of both panels (a) and (b). Poisson ratio is set to $\nu=0.35$, so that $c_D=2.082 c_S$ and $c_R=0.935 c_S$.}
\label{Fig:Afull}
\end{figure}

Figure \ref{Fig:Afull}(a)  shows the curves $A^*\:vs.\: t/\lambda_z$ obtained at various fracture speed $\vcf$. All curves can be superimposed by making $A^* \rightarrow (A^*-A^*_\infty)/(A^*_0-A^*_\infty)$ and $t \rightarrow \ccfw t /\lambda_z$ [Fig. \ref{Fig:Afull}(b)]. Thus:

\begin{equation}
\f{A^* - A^*_\infty}{A^*_0 - A^*_\infty} = g_A\left( u=\f{\ccfw t}{\lambda_z} \right).
\label{Eq:Afull}
\end{equation}

In summary, the space-time evolution of speed fluctuations for a crack propagating in a periodically patterned toughness map given by Eq. \ref{Eq:dGz} is given by:

\begin{subequations}		
  \begin{align}
 v(z,t) = & -\gamma_0 A^*(u)\sin\paf{2\pi z}{\lambda_z}\sin\pa{ \f{2\pi}{\lambda_z} c(u) t+\f{\pi}{2}},\\
 \mathrm{with} \quad & u = \f{c_{FW} t}{\lambda_z}, \\
            &  c(u|\vcf) = \ccfw^0 + (\ccfw-\ccfw^0) g_c(u), \\
               & A^*(u) = A^*_\infty + (A^*_0-A^*_\infty) g_A(u),
  \end{align}
\label{Eq:vzt_global}
\end{subequations}
 
\noindent Where the generic dimensionless functions $g_A(u)$ and $g_c(u)$ are respectively plotted in Figs. \ref{Fig:cfull}(b) and \ref{Fig:Afull}(b). The parameters $\ccfw^0$ (Eq. \ref{Eq:c0}), $\ccfw$ (pole of $p(u)$ in Eq. \ref{Eq:Pkw}), $A^*_0=1/|C_v(\vcf)|$ and $A^*_\infty$ [Eq. \ref{Eq:Ass}] are all functions of $\vcf$. The  evolutions are plotted in Figs. \ref{Fig:c0_CFW} and \ref{Fig:Ass}, respectively. Note that the cosine term $\cos(2\pi c t/\lambda_z)$ in Eq. \ref{Eq:ST_vansatz} has been replaced by $\sin(2\pi c t/\lambda_z+\pi/2)$ to be consistent with the long-time limit [Eq. \ref{Eq:vssfinal}].


\section{Dynamic crack growth perturbed by a single strip}\label{Sec4} 

In Secs \ref{Sec2} and \ref{Sec3}, we examined crack propagation along a periodically patterned toughness map. The analysis can be extended to any map as long as it remains invariant upon translation along $x$: $\gamma(x,z)=\gamma(z)$. In particular, calling $\hat{\gamma}(k)$ the $z$-Fourier transform of $\gamma(z)$, local front speed distortions can be written in $(k,t)$ space as:

\begin{equation}
\begin{split}
&\hat{v}(k,t)=-sgn(k)\hat{\gamma}(k)A^*(u)\sin\pa{c(u) k t+\f{\pi}{2}}\:\mathrm{for}\:k \neq 0,\\    
&\hat{v}(0,t)=\hat{\gamma}(0)/C_v \:\mathrm{for}\: k = 0
\end{split}
\label{Eq:vkt_global}
\end{equation}

\noindent In this expression, $u=\ccfw k t$. Note, the case $k=0$ in Eq. \ref{Eq:vkt_global} is now explicitly considered. This term was not considered since sinusoidal toughness landscapes were examined till now ($\hat{\gamma}(k=0)=0$ in this case). However, in the general case, $\gamma(k=0)\neq 0$ and, hence, $v(k=0,t)$ is to be considered. Its expression in Eq. \ref{Eq:vkt_global} was obtained by making $k=0$ in Eq. \ref{Eq:Motion_kt_final}. 

In the general case, the inverse Fourier transform of Eq. \ref{Eq:vkt_global} does not yield a simple expression for $v (z, t)$. This is due to the variations of $c$ and $A^*$ with $u$. On the other hand in the long-time limit regime, an expression is feasible. Indeed, $\hat{v}_\infty(k,t)$ writes:

\begin{equation}
\begin{split}
& \hat{v}_\infty(k,t)=A_\infty^* H(k,t)\hat{\gamma}(k) -A^*_0\hat{\gamma}(0)\delta(k), \\
& \mathrm{with} \quad H(k,t) = - sgn(k) \sin(\ccfw k t).
\end{split}
\label{Eq:vsskt_global}
\end{equation}

\noindent The inverse $z$-Fourier transform of $H(k,t)$ is: 

\begin{equation}
H(z,t)=\f{1}{2\pi}\pa{\f{1}{z-\ccfw t}-\f{1}{z+\ccfw t}}.
\label{Eq:tf_steady}
\end{equation}

\noindent Then, long-time limit velocity fluctuations become: 

\begin{equation}
\begin{split}
 & v_{\infty}(z,t) = -A_0^* \,\overline{\gamma} + F_\infty(z-\ccfw t) - F_\infty(z+\ccfw t),\\
 & \mathrm{with} \quad F_{\infty}(z) = \f{A_\infty^*}{2\pi}\int_{-\infty}^{\infty}\f{\gamma(z')-\overline{\gamma}}{z-z'}\dd z'
\label{Eq:vsszt_single}
  \end{split}
\end{equation}
\noindent where $A_0^*(\vcf)$, $A_\infty^*(\vcf)$ and $\ccfw(\vcf)$ are recalled to be function of $\vcf$ only. Here, $\overline{\gamma}$ is the average of $\gamma(z)$ over $z$.

Of particular interest is the situation with a single strip localized at a position $z_0$, of width $\xi_z$, amplitude $\gamma_0$, and dimensionless shape $\gamma^*(z^*)$ defined so that:

\begin{equation}
\f{\gamma(z)}{\gamma_0}= \gamma^*\left(z^*=\f{z-z_0}{\xi_z}\right).
\label{Eq:gammastrip}
\end{equation}

\begin{figure}
\includegraphics[width=\columnwidth]{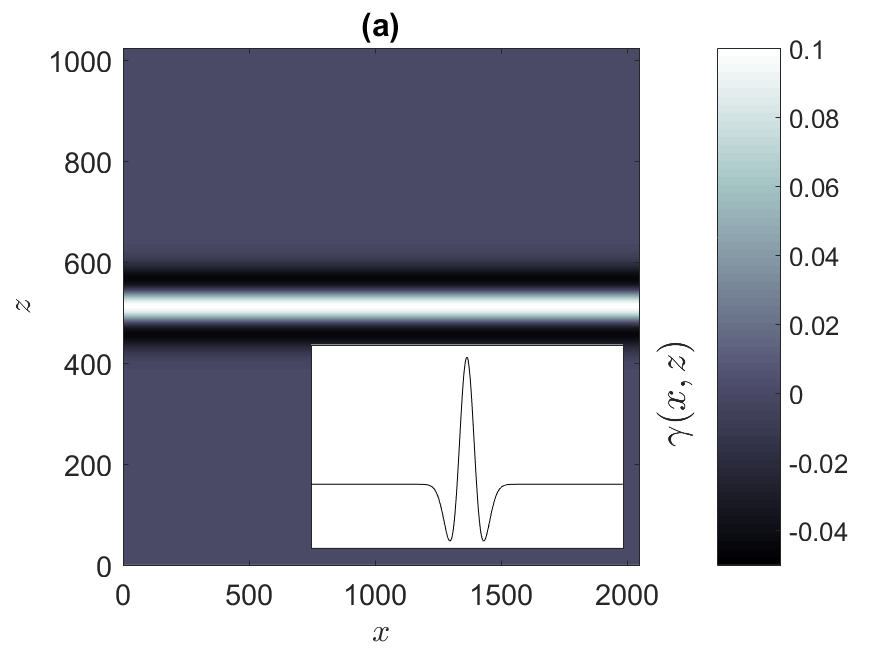}\\
\includegraphics[width=\columnwidth]{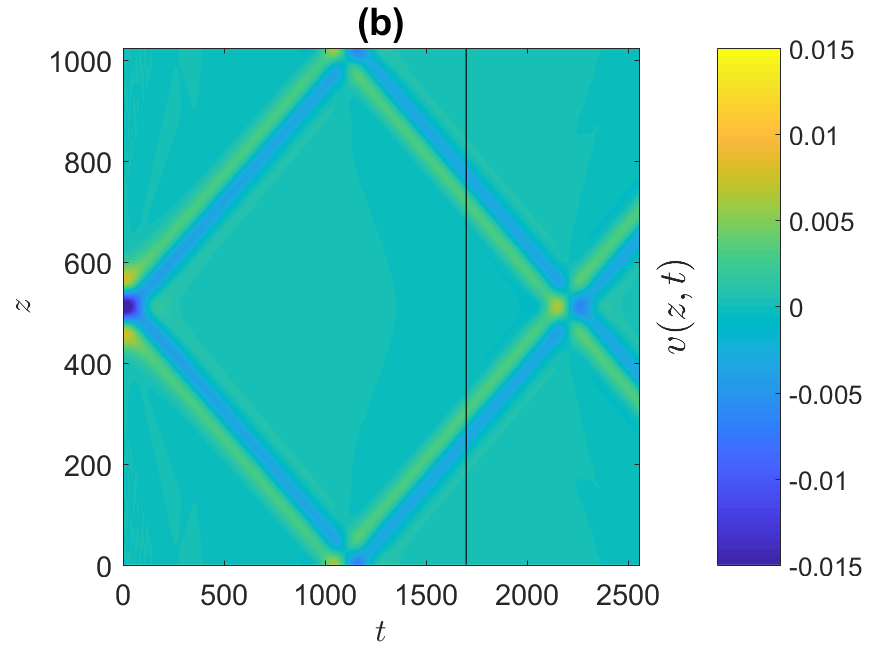}
\includegraphics[width=\columnwidth]{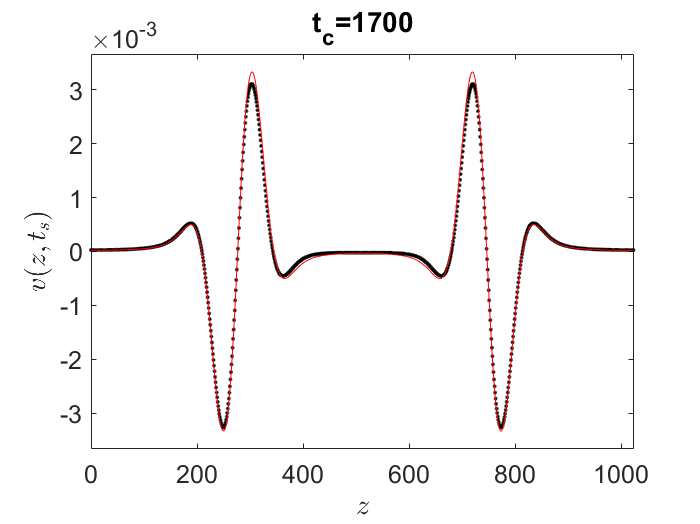}
\caption{(a) Toughness map with a localized strip of Mexican hat shape (inset): $\gamma^*(z^*)=(1-{z^*}^2)\exp(-z^*/2)$. Strip amplitude and size are $\gamma_0=0.1$ and $\xi_z=32$. Map size is $L=1024$. (b) Space-time evolution of the local velocity $v(z,t)$ for a crack propagating in the landscape shown in panel (a). Black vertical line indicates time $t_c=1700$ when profile $v(z,t_c)$ is recorded. (c) Variation of local velocity $v(z,t_c)$ as a function of $z$ at time $t_c$. Black dots show $v(z,t_c)$ as obtained from the numerical simulation, while red solid line is the solution provided by Eq. \ref{Eq:vsszt_single}. In panels (b) and (c) $\vcf=0.8 c_S$. Poisson ratio is set to $\nu=0.35$ so that $c_D=2.0832 c_S$ and $c_R=0.935 c_S$.}
\label{Fig:map_single}
\end{figure}

\noindent Such toughness maps were constructed [Fig. \ref{Fig:map_single}(a)] and the spatio-temporal evolution of the local speed was determined using the numerical scheme described in Sec. II.D [Fig. \ref{Fig:map_single}(b)]. Once moving front begins to interact with this strip, two localized disturbances form in $v(z,t=0)$. In the long-time limit, these two disturbances are of opposite signs (see minus sign in between the two right-hand terms of first Eq. \ref{Eq:vsszt_single}) and propagate upwards and downwards along the front, with the FW speed $\ccfw$. In the long-time limit (i.e. for $t\gg \xi_z/\ccfw$), the profile of velocity fluctuations as determined numerically is in very good agreement with Eq. \ref{Eq:vsszt_single} [Fig. \ref{Fig:map_single}(c)].   

By injecting Eq. \ref{Eq:gammastrip} into Eq. \ref{Eq:vsszt_single}, it is possible to express the long-time limit pulse shape, $F_{\infty}(z)$, in a dimensionless form:

\begin{equation}
\f{F_\infty(z)}{A^*(\vcf)\gamma_0} = F_{\infty\gamma^*}^*\left(z^*=\f{z-z_0}{\xi_z}\right).
\label{Eq:pulsenorm1}
\end{equation}

\noindent This form is generic, independent of strip amplitude, width and position; it is selected by the strip shape only:  

\begin{equation}
F_{\infty\gamma^*}^*(z^*) = \f{1}{2\pi}\int_{-\infty}^{\infty}\f{\gamma^*({z^*}')-\overline{\gamma^*}}{z^*-{z^*}'}\dd {z^*}'
\label{Eq:pulsenorm2}
\end{equation}

\begin{figure}
\includegraphics[width=\linewidth]{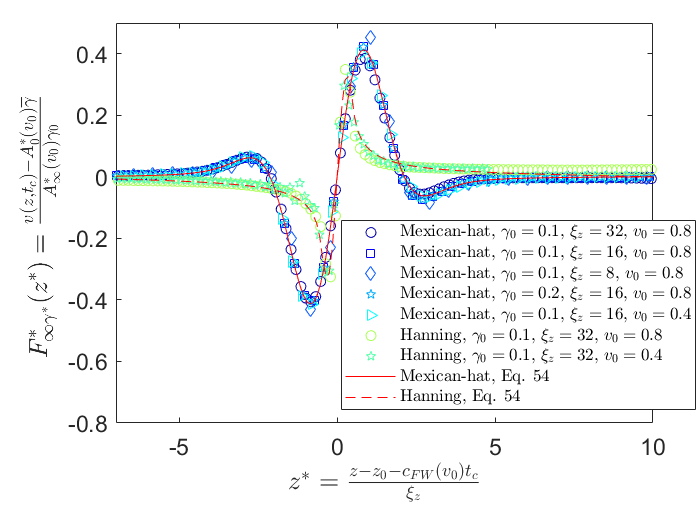}
\caption{Generic shape for the FW pulse generated when a dynamic crack propagates at $\vcf$ along a toughness with a single strip of Mexican hat profile ($\gamma^*(z^*)=(1-{z^*}^2)\exp(-{z^*}^2/2)$) or Hanning profile ($\gamma^*(z^*)=\cos(\pi z^*/2)$ for $|z^*|<1/2$, $\gamma^*(z^*)=0$ for $|z^*|\geq 1/2$) and different parameters $\gamma_0$ and $\xi_z$. The different data point symbols correspond to the velocity profiles along $z$ at $t_c=1700$ obtained in the simulations achieved with the parameters provided in the legend, once rescaled via Eq. \ref{Eq:pulsenorm1}. Red solid line and red dashed line are the analytical predictions provided by Eq. \ref{Eq:pulsenorm2} for the Mexican-hat strip shape and the Hanning strip shape, respectively.}
\label{Fig:norm-pulse}
\end{figure}

\noindent Figure \ref{Fig:norm-pulse} test this prediction against numerical simulations performed at two $\vcf$, different parameters $\gamma_0$ and $\xi_z$, and two different strip shapes $\gamma^*(z^*)$: Mexican hat shape ($\gamma^*(z^*)=(1-{z^*}^2)\exp(-{z^*}^2/2)$) and Hanning shape ($\gamma^*(z^*)=\cos(\pi z^*/2)$ for $|z^*|<1/2$, $\gamma^*(z^*)=0$ for $|z^*|\geq 1/2$). The scaling proposed in Eq. \ref{Eq:pulsenorm1} is fulfilled very well, and the analytical predictions given in Eq. \ref{Eq:pulsenorm2} are also fullfilled. 

\section{Concluding discussion}\label{Sec5}

We have examined numerically and analytically the propagation of a dynamic crack along a 2D plane made of unidimensional strips with periodically modulated toughness. The space-time structure of local speed fluctuations was found to exhibit the check-board structure characteristic of standing FW. An analytical solution was obtained in the long-time limit, and an asymptotic expansion up to any order was obtained for the initial transient regime. These solutions were finally shown to extend to any toughness landscape provided it remains invariant along the mean direction of crack propagation.

Several important features emerge from this analysis. First, there is a clear initial regime where both FW amplitude and speed are different from what they are in the long-time limit regime. The typical duration of this transient is set by the ratio between the toughness in-plane length-scale (sinusoidal wavelength or strip width), and the FW speed. FW amplitude starts at a value about two times larger than long-time limit value, then decreases and saturates to this long-time limit value.  It is of interest to note that both the short-time and long-time limit regimes obey a simple d'Alembert 1D wave equation with different wave speeds and source functions; this may enable cost-effective numerical simulations of both regimes in more complex situations. This is consistent with what is observed in numerical simulations of dynamically growing crack interacting with a single localized asperity \cite{Morrissey00_jmps,Fekak20_jmps}. This FW speed also evolves in the transient regime: it starts at a value about two times less than the long-time limit value. Then increases to reach the long-time limit value. This may explain why the FW speed measured in simulations (single asperity configuration) \cite{Fekak20_jmps} are slightly (but noticeably) smaller than the value predicted theoretically, from the zero of the elastodynamic kernel  $P(k,\omega)$ (Eq. \ref{Eq:Pkw}). Finally, it is interesting to note that FW pulses emitted as a dynamic front interacts with a single strip takes a generic dimensionless scaling form (Eq. \ref{Eq:pulsenorm1}).

Obtaining simple exact analytical results was made possible here since the examined configuration is simple: 2D planar crack with toughness landscape invariant along crack growth direction. However, most situations of experimental and engineering interests consider cracks interacting with one or several localized inclusions (e.g. placed to reinforce material \cite{Becher91_jacs}). It is therefore of interest to extend the analysis derived here to this situation. The problem of how a crack front roughens as it moves through a spatially-distributed toughness landscape has also been widely considered within the elastostatic limit \cite{Schmittbuhl95_prl,Ramanathan97_prl,Chopin15_jmps} and our analysis may be used to incorporate the elastodynamics effects to these approaches. Works in these directions are currently in progress. Let us finally mention that, beyond Eq. \ref{Eq:Kperturbed}, there exists solutions \cite{Willis97_jmps,Willis12_jmps,AddaBedia13_prl} providing the relations between the local variation of all modes of stress intensity factors and both in-plane and out-of-plane front distortions. Hence, it may be possible to extend this work to shear fracture and/or full 3D problem, with predictions on the out-of-plane roughness in which experiments have sought so far signatures of FW  \cite{Sharon01_nature,Wang20_pnas}. Work in this direction is also in progress.


\appendix
\section{Series solution up to any order $m$}\label{A1}

First, let us examine the short time limit. The idea to solve Eq. \ref{Eq:Motion_kt_final} is to write $\hat{\base}(u=k t |\vcf)$ as a power series and seek solutions $\hat{f}(k,t)$ of the form:

 \begin{equation}
 \hat{f}(k,t)=\Sum{m=1}{\infty} {f_m(k)\,t^m}
 \label{Eq:fs1}
\end{equation}

\noindent Note that the zeroth order term $f_0=0$; since $f(t=0)=0$ is prescribed as an initial condition. The Bessel function of the first kind $J_\nu(u)$ writes:

 \begin{equation}
 J_\nu(u)=\left(\frac{u}{2}\right)^{\nu}\Sum{m=0}{\infty} \mathlarger{\f{(-1/4)^m}{m!(\nu+m)!}}u^{2 m}
\end{equation}

\noindent By replacing $J_0(u)$, $J_1(u)$ and $J_2(u)$ by the above series in Eq. \ref{Eq:BB}, we get:

\begin{equation}
\begin{split}
\hat{\base}(u|\vcf) &= \Sum{m=0}{\infty} B_m(\vcf) u^{2 m} \quad \mathrm{with:} \\
  B_m(\vcf) & =  \f{(-1/4)^m}{m!\,(m+1)!}\left(\f{1}{2} c_D(\alpha_D c_D)^{2m} - c_R(\alpha_R c_R)^{2m} \right.\\
 & \left. -\f{1}{2}\Int{c_S}{c_D}\Theta(\eta)\left[\f{\eta^2+\vcf^2}{\eta^2-\vcf^2}m+(m+1)\right](\alpha_\eta \eta)^{2m}\,\dd\eta \right)
\end{split}
\label{Eq:BBs1}
\end{equation}

\begin{figure}
\includegraphics[width=\linewidth]{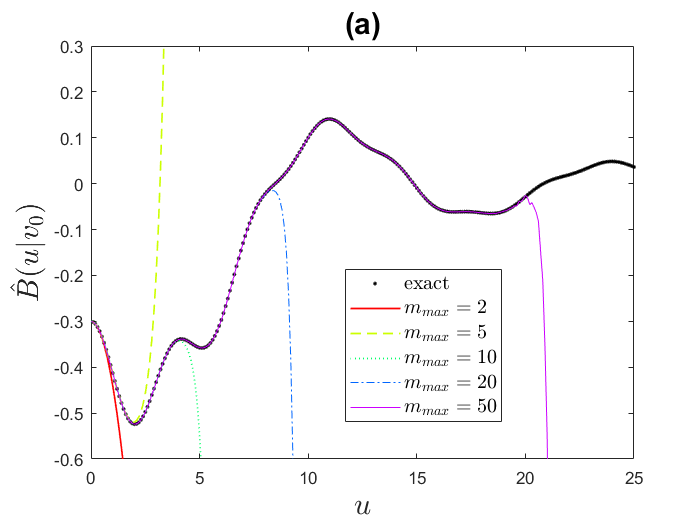}
\includegraphics[width=\linewidth]{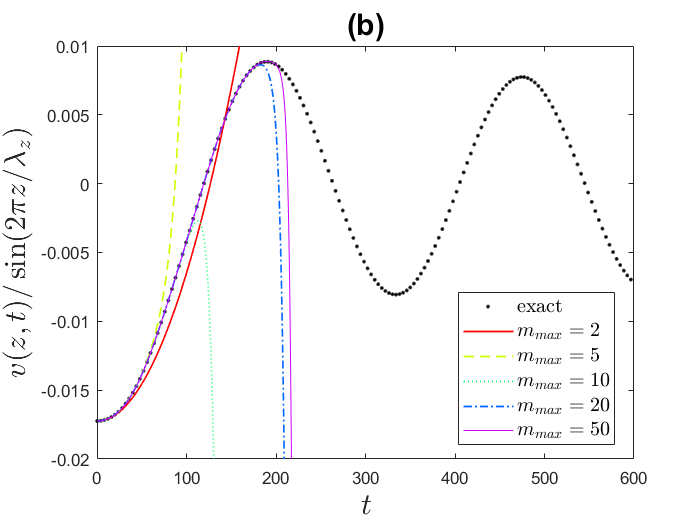}
\caption{(a) Comparison between $\hat{B}(u|\vcf)$ as computed directly using Eq. \ref{Eq:BB} (black points), and as approximated using the series expansion given by Eq. \ref{Eq:BBs1}. The various colored lines correspond to increasing truncations $m_{max}$: $m_{max}=2$, $m_{max}=5$, $m_{max}=10$, $m_{max}=20$ and $m_{max}=50$. Here, $\vcf=0.8$ and the black point curve (numerical $\hat{B}(u|\vcf)$) is the same as that represented in Fig. \ref{Fig:cvv_and_BI}(b), zoomed on shorter $u$. (b) Comparison between $v(z,t)/\sin(2\pi z/\lambda_z)$ as obtained by simulations (black points), and as obtained analytically using the series expansion given in Eq. \ref{Eq:vs1} at increasing order $2m$. Here, $\vcf=0.8 c_S$, $\gamma_0=0.1$, $N=1024$ and $\lambda_z=128$; the black point curve (numerical $v(z,t)/\sin(2\pi z/\lambda_z)$) is the same as that represented in Fig. \ref{Fig:scalingv_per_z}, zoomed on shorter $t$.}.
\label{Fig:DL-v_vs_t}
\end{figure}

\noindent As shown in Fig. \ref{Fig:DL-v_vs_t}(a), this series expansion approximates $\hat{B}(u|\vcf)$ very well at short times.

Introducing this series expansion and Eq. \ref{Eq:fs1} in Eq. \ref{Eq:Motion_kt_final} yields:

\begin{equation}
\begin{split}
& C_v(\vcf) \Sum{m=0}{\infty} {(m+1)\,f_{m+1}\,t^{m}} \\
+ & k^2 \int_{0}^t \Sum{m=0}{\infty} B_{m} (k(t-t'))^{2m} \Sum{n=0}{\infty} f_n(k) t'^n \dd t' =
                                    \hat{\gamma}(k) 
\end{split}
\end{equation}

\noindent After reversing integral and the sum in the second left-hand term, one gets:

\begin{equation}
\begin{split}
& C_v(\vcf) f_1 + \Sum{m=1}{\infty} {C_v(\vcf) (m+1)f_{m+1} t^{m}} \\
+ & k^2 \Sum{m=1}{\infty} \pa{\Sum{n=1}{m}B_{m-n} f_{2n-1} g_{m-n,2n-1}}t^{2m} \\
+ & k^2 \Sum{m=0}{\infty} \pa{\Sum{n=0}{m}B_{m-n} f_{2n} g_{m-n,2n}}t^{2m+1}  =
                                    \hat{\gamma}(k). \end{split}
\end{equation}

\noindent where $g_{m,n}$ is defined by:

\begin{equation}
g(m,n)= \Sum{p=0}{m} \binom{m}{p} \f{(-1)^p}{p+n+1}
\end{equation}

\noindent By balancing the successive coefficients in front of each $t^m$ term, we determine the coefficients $f_m(k)$: 
\begin{equation}
\begin{split}
& f_0 = 0, \\
& f_1 = \hat{\gamma}(k) / C_v(\vcf), \\
& f_{2m} = 0,  \\
& f_{2m+1} = \f{-k^2}{C_v(\vcf)(2m+1)} \\
& \quad\quad\quad \times \Sum{n=0}{m-1}B_{m-1-n} k^{m-1-n} g(m-1-n,2n+1)f_{2n+1}.
\end{split}
\end{equation}

\noindent Differentiating Eq. \ref{Eq:fs1} with respect to time yields:

\begin{equation}
\begin{split}
& \hat{v}(k,t)=\f{\hat{\gamma}(k)}{C_v(\vcf)}+\Sum{m=1}{\infty} {v_{2m}(k t)^{2m}} \quad \mathrm{with:}\\
& v_{2m} =\f{-1}{C_v(\vcf)}\Sum{n=0}{m-1}B_{m-1-n}g(m-n-1,2n+1)f_{2n+1}
\end{split}
\label{Eq:vs1}
\end{equation}

\noindent Equation \ref{Eq:vs1} provides a solution up to any order $2m$. Consider now the sinusoidal toughness given by Eq. \ref{Eq:dGz}, its $z$-Fourier transform yields:

\begin{equation}
\hat\gamma(k)=i\pi\gamma_0\left[\delta\pa{k+\f{2\pi}{\lambda_z}} - \delta\pa{k-\f{2\pi}{\lambda_z}} \right]
\label{Eq:dGk}
\end{equation}

\noindent By replacing $\hat{\gamma}(k)$ in Eq. \ref{Eq:vs1} by its expression above, and then by applying inverse $z$-Fourier transform, we get:

\begin{equation}
\begin{split}
& v(z,t)=\f{\gamma_0}{C_v(\vcf)}\sin\paf{2\pi z}{\lambda_z}+\Sum{m=1}{\infty} {v_{2m}\paf{2\pi t}{\lambda_z}^{2m}} \quad \mathrm{with:}\\
& v_{2m} =\f{-1}{C_v(\vcf)}\Sum{n=0}{m-1}B_{m-1-n}g(m-n-1,2n+1)f_{2n+1}
\end{split}
\label{Eq:vsfinal}
\end{equation}

\noindent As shown in Fig. \ref{Fig:DL-v_vs_t}(b), this analytical solution coincides perfectly with the numerical result at short times.

\section{Evaluation of the integral involved in the inverse Fourier transform for the long-time limit solution}\label{B1}

\begin{figure}
\includegraphics[width=\linewidth]{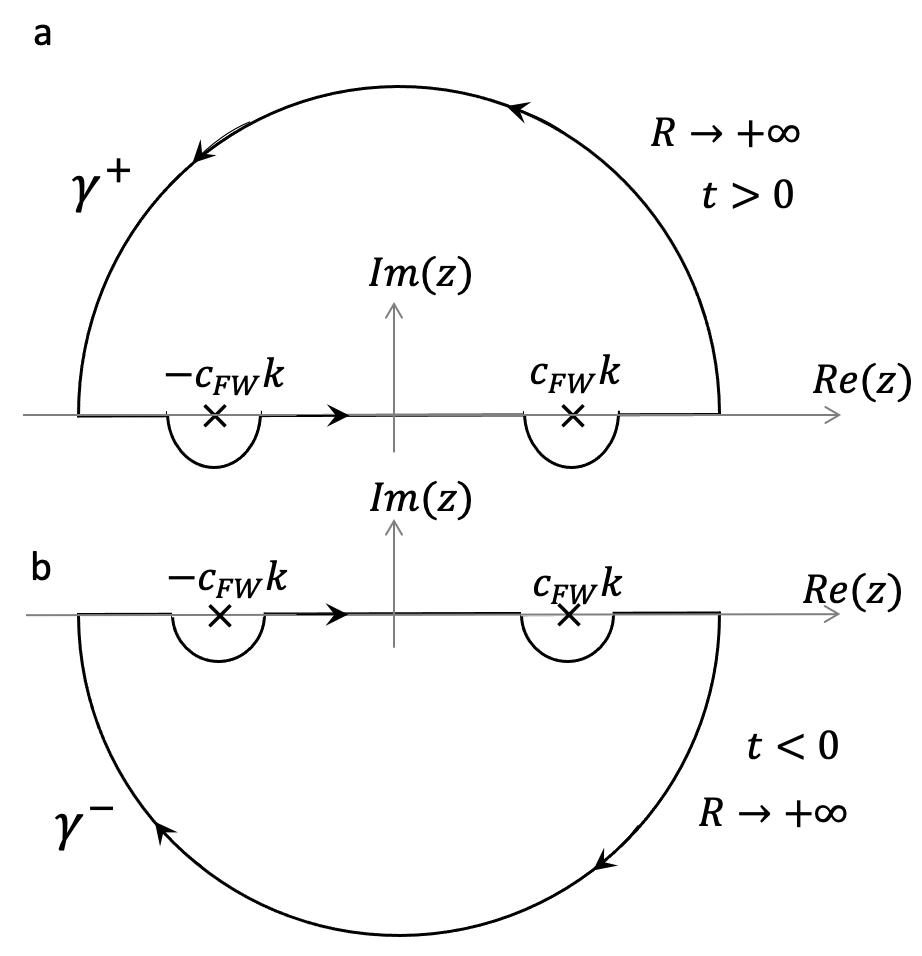}
\caption{Closed contour along which $\exp(izt)/(z^2-k^2\ccfw^2)$ is integrated. (a) Contour $\gamma^+$ used for the positive times; (b) Contour $\gamma^-$ used for negative times.}
\label{Fig:residue}
\end{figure}

As explained in Sec. \ref{Sec:longtime}, the long-time limit solution $v_\infty(k,t)$ for the velocity fluctuations is given by Eq. \ref{Eq:v_kt_infty}, that is:

  \begin{equation}
  \hat{v}_\infty(k,t) = -\f{\hat{\gamma}(k)|k|}{2\pi \,p'(c_{FW}^2)}\int_{-\infty}^\infty \f{\exp(i\omega t)}{\pa{\omega^2-k^2\,c_{FW}^2}} \dd \omega
\label{vss_kt}
 \end{equation}

 \noindent The integrand has two simple poles $\omega_\pm = \pm \ccfw k$. The associated residues are given by:

\begin{equation}
Res(\omega_\pm)=\lim\limits_{\omega\to \omega_\pm}\f{(\omega-\omega_\pm)\exp(i \omega t)}{\pa{\omega^2-k^2\,c_{FW}^2}}
\end{equation}

\noindent This yields $Res(\omega_\pm)=\pm e^{\pm i \ccfw k t}/(2\ccfw k)$. To compute the integrand in \ref{vss_kt}, we use complex analysis and the residue theorem (see e.g. \cite{Appel07_book}, Sec. 15.3 for similar analysis). Let us consider the integration of the function $\exp(izt)/(z^2-k^2\ccfw^2)$ along two closed contours in the complex plane $z$: $\gamma^+$ for $t>0$ [Fig. \ref{Fig:residue}(a)], and $\gamma^-$ for $t<0$ [Fig. \ref{Fig:residue}(b)]. The second Jordan lemma tells us that, for $t\geq 0$ : 

  \begin{equation}
 \int_{-\infty}^\infty \f{\exp(i\omega t) }{\pa{\omega^2-k^2\,c_{FW}^2}}\dd \omega =\int_{\gamma^+} \f{\exp(i\,z \,t)}{\pa{z^2-k^2\,c_{FW}^2}} \dd z,
\label{vss_kt1+}
 \end{equation}
 
\noindent and for $t<0$: 

  \begin{equation}
 \int_{-\infty}^\infty \f{\exp(i\omega t) }{\pa{\omega^2-k^2\,c_{FW}^2}}\dd \omega =\int_{\gamma^-} \f{\exp(i\,z \,t)}{\pa{z^2-k^2\,c_{FW}^2}} \dd z
\label{vss_kt1-}
 \end{equation}
 
\noindent Then, the application of the residue theorem yields: 

  \begin{equation}
\int_{\gamma^+} \f{\exp(i z t)\dd z}{\pa{z^2-k^2 c_{FW}^2}} =2 i\pi\left(Res(\omega_+)+Res(\omega_-)\right)
\label{vss_kt2+}
 \end{equation}

\noindent and:

  \begin{equation}
\int_{\gamma^-} \f{\exp(i\,z \,t)\dd z}{\pa{z^2-k^2\,c_{FW}^2}} =0
\label{vss_kt2-}
 \end{equation}
 
\noindent As a result, we get:

\begin{equation}
\begin{split}
& \mathrm{for}\, t < 0,\,\hat{v}_{\infty}(k,t)=0\\
& \mathrm{for}\, t\geq 0, \hat{v}_{\infty}(k,t)=\f{sgn(k)\hat{\gamma}(k)}{\ccfw p'(\ccfw^2)}\sin(\ccfw k t)
\end{split}
\label{Eq:vss_ktf}
\end{equation}

\noindent In the following, we only consider $t\geq 0$. We use Eq. \ref{Eq:dGk} to go back into the real $(z,t)$ space:  

\begin{multline}
\hat{v}_{\infty}(z,t)=\f{i \gamma_0}{2\ccfw p'(\ccfw^2)}\int_{-\infty}^{\infty} 
sgn(k) \sin(\ccfw k t) \times \\[5mm]
\left[\delta\pa{k+\f{2\pi}{\lambda_z}}-\delta\pa{k-\f{2\pi}{\lambda_z}}\right] \exp(i k z)\dd k
\end{multline}

\noindent Subsequently $v_\infty(z,t)$ takes the form given by Eq. \ref{Eq:vssfinal}. 

\section*{Acknowledgments}

We thank Mokhtar Adda-Bedia, Fabian Barras, Eric Brillaux, Fatima Fekak, Jean-Francois Molinari and Henning Samtleben for fruitful discussions. We also thank Cindy Rountree for the careful reading of the article. Funding by "Investissements d’Avenir" LabEx PALM (ANR-10-LABX-0039-PALM) is also gratefully acknowledged.


%

\end{document}